\documentclass[11pt]{article}
\usepackage{epsfig}
\usepackage{adjustbox}
\usepackage{diagbox}
\usepackage{dcolumn}
\usepackage{bm}
\usepackage{graphicx}
\usepackage[textsize=tiny]{todonotes}
\usepackage{subcaption}
\usepackage{here}
\usepackage{amssymb,amsmath}
\usepackage{multirow}
\usepackage{cite,color,url}
\usepackage{tabu}
\usepackage{array}
\usepackage[colorlinks=true,urlcolor=blue,anchorcolor=blue
,citecolor=blue,filecolor=blue,linkcolor=blue,menucolor=blue
,linktocpage=true,pdfproducer=medialab,pdfa=true]{hyperref}

\usepackage{colordvi}
\usepackage{epsfig,psfrag,rotating,soul}
\def\beqa{\begin{eqnarray}}
\def\eeqa{\end{eqnarray}}

\evensidemargin \oddsidemargin
\allowdisplaybreaks
\renewcommand{\arraystretch}{1.5}
\psfrag{lk}{$l_k$}
\psfrag{lm}{$l_m$}
\psfrag{blk}{$\bar {l}_k$}
\psfrag{blm}{$\bar {l}_m$}
\def\thefootnote{\fnsymbol{footnote}}

\let\OLDthebibliography\thebibliography
\renewcommand\thebibliography[1]{
\OLDthebibliography{#1}
\setlength{\parskip}{0pt}
\setlength{\itemsep}{0pt plus 0.3ex}}
\usepackage{makecell}

\def\tr{\text{Tr}}
\def\zt{\mathbb{Z}_2}
\def\thdm{2{\text HDM}}
\def\idm{{\text IDM}\,}

\def\htm{{\text HTM}\,}

\def\gev{{\text GeV}}

\begin{document}

\thispagestyle{empty}
\begin{center}
\begin{Large}
\textbf{\textsc{Higgs-like particle decays into $\gamma Z$ and $\gamma\gamma$ : Fingerprints of some non-supersymmetric models.}}
\end{Large}

\vspace{1cm}
{
R.Benbrik$^{1}$%
\footnote{\tt \href{mailto:r.benbrik@uca.ma}{r.benbrik@uca.ma}}
M. Boukidi,$^{1}$%
\footnote{\tt
\href{mailto:mohammedboukidi@gmail.com}{mohammedboukidi@gmail.com}}
M. Ouchemhou,$^{1}$%
\footnote{\tt
\href{mailto:mohamed.ouchemhou@ced.uca.ac.ma}{mohamed.ouchemhou@ced.uca.ac.ma}}
L. Rahili,$^{2}$%
\footnote{\tt
	\href{mailto:rahililarbi@gmail.com}{rahililarbi@gmail.com}}
} and
O. Tibssirte$^{3}$%
\footnote{\tt
\href{mailto:oumayma1tibssirte@gmail.com}{oumayma1tibssirte@gmail.com}}

\vspace*{.7cm}
{\sl
$^1$ Polydisciplinary Faculty, Laboratory of Fundamental and Applied Physics,
Cadi Ayyad University, Sidi Bouzid, B.P. 4162, Safi, Morocco}\\
{\sl
$^{2}$ Laboratory of Theoretical and High Energy Physics (LPTHE), Faculty of Science, Ibn Zohr University, B.P 8106, Agadir, Morocco.}\\
{\sl
$^{3}$ UPM University, B.P 42312, Marrakech, Morocco.}
\end{center}

\vspace*{0.1cm}

\begin{abstract}
Recently, ATLAS and CMS experiments at the LHC put on light the relevant results in the measurement precision of the Higgs and BSM. In such a report, where the resonance direct search was made in the $\gamma Z$ channel, a mass adjustment distribution for the reconstructed $Z$ boson and photon was established. Thus, simultaneously with the signal-from-background separation, the number of events has been perfectly described, and an excess signal approximately twice that expected by the Standard Model (SM) has been noticed, which is equivalent to a significance of 2.2 standard deviations. 
In this study, we examine how any possible new physics models can explain this excess, such as the CP-conserving Two-Higgs doublet model (2HDMs), the Inert doublet model (IDM), and the Higgs triplet model (HTM). While considering the available theoretical and most recent experimental constraints, the phenomenological implications of existing extensions beyond the SM are discussed, and prospects for precision studies of these processes are described. We have found that the excess could be explained in the BSM frameworks studied,  depending on the charged and double-charged Higgs boson masses.
\end{abstract}
 
\def\thefootnote{\arabic{footnote}}
\setcounter{page}{0}
\setcounter{footnote}{0}
\section{Introduction}
\label{sec:intro}
The Large Hadron Collider (LHC) machine declared the success of its first run in the CMS and ATLAS by discovering a new scalar particle\cite{Aad:2012tfa,Chatrchyan:2012ufa} consistent with the Higgs boson predicted by the SM. The discovery of this particle was with a resonance of around 125 GeV, and the measurement of its properties \cite{ATLAS:2013dos,CMS:2013btf,ATLAS:2013xga,CMS:2012vby} has been considered one important challenge to assess this responsibility. In this energy range, the SM is regarded as the fundamental theory of physics. Nevertheless, with the rise of some unusual signatures and unexplained deviations compared to SM, it's obvious that more fundamental theories should appear at higher energies, preferably on the TeV scale, allowing the LHC to test such new theories. One of the sectors that provide wide and rich phenomenological diversities, the Higgs sector seems to be the most involved one beyond the Standard Model (BSM) due to its nature, which is not yet determined. Furthermore, the corresponding observables ought to be calculated more precisely to distinguish various Higgs sectors. Examples of this latter, although not exhaustive, could include the $\rho$ parameter\cite{Diaz-Cruz:2003kcx}. Consequently, one can distinguish two different categories of BSM extensions: those which uphold the $\rho=1$ as in the SM and the others where such a parameter is slightly different from the unity. Such former models with $\rho \approx 1$ are: the Two-Higgs doublet models (2HDM and IDM)\cite{Branco:2011iw,Hessenberger:2016atw} and the Higgs triplet model (HTM)\footnote{\scriptsize \textcolor{black}{It’s well known that the HTM, does not preserve tree-level custodial symmetry since the $\rho$ parameter is shifted with a negative $\Delta\rho=\alpha_{EW}\,T$, which means that at tree-level, the HTM contributes only to the $T$ parameter. In addition, as long as $\Delta\rho$ must be small, $v_{\Delta}$ should not exceed an upper limit, i.e., $v_{\Delta} \le 3$ GeV.  However, the HTM embeds additional Higgs bosons that contributes, at one-loop level, positively to $T$ and furthermore dominate over the tree-level contribution.  Within $h\equiv\,h_{SM}$ so $c_\alpha\simeq 1$, assuming $v_{\Delta}=1$, the hierarchy $m_{H^{++}} < m_{H^{+}} < m_{A} \equiv m_{H}$ ($m_{H^{++}} > m_{H^{+}} > m_{A} \equiv m_{H}$) could occur and a mass splitting around 10 to 40 GeV between the components is predicted for $|T|$ around $1\sigma\sim2\sigma$. }}\cite{Kanemura:2012rs}, which are reviewed here to explain the $\mu_{\gamma Z}$ behavior.

The ATLAS collaboration reveals a modest excess of detected events in the $H \to \gamma\gamma$ decay channel. Additional charged particles contributing to the loop that mediates the decay could explain that excess\cite{Ferreira:2012my,Ferreira:2011aa,Chiang:2012qz}. Such charged particles would participate in the same perturbative order as those of the SM, making this decay extremely sensitive to novel physics. Another comparable decay to which the new charged particles would contribute in this scenario is the $H\to \gamma Z$, although it has not been observed experimentally.  The $H\to \gamma\gamma$ and $H\to \gamma Z$ processes, although being highly suppressed decays, may thus be dominant in our understanding of plausible phenomena beyond the SM (BSM) theories.

The $H \to \gamma Z$ and $H \to \gamma\gamma$ decays occurred after the Higgs production at the LHC, following one of the standard mechanisms of production. The highest production cross section for the Higgs boson in the SM occurs via gluon fusion (ggH), which is almost an order of magnitude more powerful than vector boson fusion (VBF) and other processes in the mass range below the $H\to W^\pm W^\mp$ decay threshold. Even if some of the production channels may include extra leptons, jets, or missing energy in their final state, it will be challenging to utilize these various signatures, at least at low luminosity. As a result, the inclusive $H \to \gamma Z$ and $H \to \gamma\gamma $ processes will be our primary focus. It will also be described how undertaking specialized research, such as producing particles through vector boson fusion (VBF), is appealing since it makes it easier to identify the kinds of novel theories that can be tested in the $H\to \gamma Z$ and $H\to \gamma\gamma$ modes. However, in this situation, a large integrated luminosity is required. Similar to the decays $H\to \gamma Z$ and $H\to \gamma\gamma$, the primary manufacturing process (ggH) is a loop-induced process that is affected by the same physical principles.
The CMS \cite{CMS:2013rmy,CMS:2018myz} and ATLAS \cite{ATLAS:2014fxe,ATLAS:2017zdf} collaborations are scouring the LHC for evidence of the $H \to \gamma Z$ disintegration, leading to the most recent upper limit on the signal strength $\mu_{\gamma Z} < 3.9$ and $\mu_{\gamma Z} < 6.6$, respectively. Precision measurements for the signal strengths of these decays, $H\to \gamma Z$ and $H\to \gamma \gamma$, can achieve respective values of $\mu_{\gamma Z } = 1 \pm 0.23$ and $\mu_{\gamma\gamma} = 1 \pm 0.04$ for both ATLAS and CMS in the future project from the LHC\cite{Cepeda:2019klc} with its High Luminosity (HL-LHC) and High Energy (HE-LHC). Additionally, it is hoped that the ATLAS estimated significance to the $H \to \gamma Z$ channel will be $4.9\sigma$ with $3000$ fb$^{-1}$\cite{Cepeda:2019klc,ATLAS:2018jlh}. 

Following this disclosure by ATLAS and CMS of potential excess in the diphoton and $Z-$photon channel, numerous studies addressed  this improvement in a variety of models, including SUSY and extended Higgs sector models such as the inert, and triplet Higgs models\cite{Arhrib:2012ia,Swiezewska:2012eh,Arhrib:2012vp,Akeroyd:2012ms,Chun:2012jw,Kanemura:2012rs,Melfo:2011nx,Wang:2013jba,Blunier:2016peh}, the 2HDM\cite{Fontes:2014xva,Kanemura:2018yai,Bhattacharyya:2013rya,Bhattacharyya:2014oka,Ferreira:2012my,Ferreira:2011aa,Altmannshofer:2012ar,Fontes:2014tga}, MSSM\cite{Kitahara:2012pb}, the model 331\cite{Hung:2019jue}, the scotogenic model\cite{Chen:2019okl} and, others in the BSM frameworks \cite{Delgado:2012sm,Chiang:2012qz}. Furthermore, there are many studies in the literature focusing on the $H\to \gamma Z$ decay and its correlation to $H \to \gamma\gamma$\cite{Chen:2013dh,Chen:2013vi,BhupalDev:2013xol,Chiang:2012qz}. 

As for the HL-LHC which is an upgraded version of the LHC that will increase the luminosity, or the amount of particle collisions per second, by up to one order of magnitude.  Given more data, the HL-LHC will be used to investigate and study charged Higgs bosons through their interaction with photons via $H\to \gamma\gamma$ and/or $H\to Z \gamma$ decay modes, the latter, has not yet been observed and still the target of theoretical and experimental investigation. With more data from the HL-LHC will help test the predictions of the Standard Model and look for deviations, including the possibility of setting new bounds on  Z photon channel. The limits on the mass of a charged Higgs boson, however, depend on the specific model considered. With the increased luminosity provided by the HL-LHC, these limits should be improved considerably, which could lead to the discovery of a charged Higgs boson or to new constraints on its existence during the Run-III.

In this work, we review the prediction of the signal strengths $\mu_{\gamma Z}$ and, $\mu_{\gamma \gamma}$ in the theories outside the realm of the standard model, such as the Two-Higgs Doublets model (2HDM) as well as its special case, the Inert Doublet Model (IDM), and the Higgs Triplets Model (HTM), bearing in mind the most current and available theoretical and experimental restrictions.

This paper is arranged in a way where we start with the $\gamma Z$ SM decay perspective in Section \ref{sec:hgammaZ-sm}, followed by a brief introduction of BSM extensions and the available theoretical and experimental constraints in Section \ref{sec:hgammaZ-bsm}, then we exhibit the results and discussion in Section \ref{sec:res}, and we come to conclude in Section \ref{sec:Conclusion}.
\section{The Higgs-photon-$Z$ boson vertex in the SM}
\label{sec:hgammaZ-sm}
\begin{figure}[H]
	\center 
	\includegraphics[scale=1]{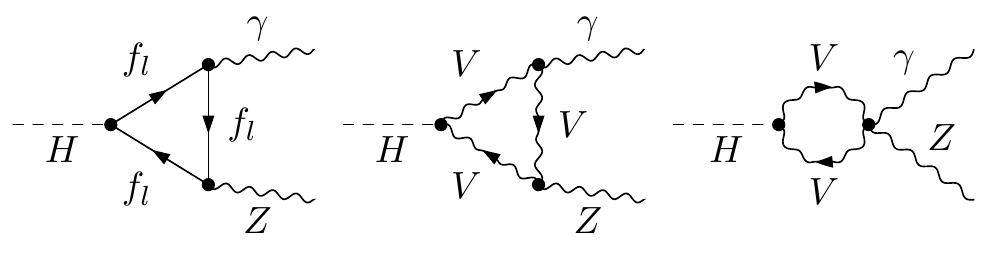}
	\caption{The generic Feynman diagrams for $H \to  \gamma Z$.}
	\label{fig:diagrams-sm}
\end{figure}

Beginning with an overview of the process $H \to \gamma Z$ within the context of the SM, and based on the fact that such a process does not occur at tree-level, we point out in Fig.\ref{fig:diagrams-sm} all massive particles that act as intermediaries between the Higgs boson and final states. These loops could be generated at the quantum level, and 
might be illustrated by the following Feynman diagrams, obtained with the help of the FeynArts \cite{Hahn:2000kx} package,
\noindent
where $f_l$ and $V$ stand for fermions and gauge boson, respectively. Including only the dominant top contribution ($f_l \equiv t$) multiplied by the color factor $N_f$, as well as the $W$ boson loops, the partial width is built up and expressed in \cite{Cahn:1978nz,Djouadi:2005gi,Spira:2016ztx} as,
\begin{equation}
\Gamma(H \to \gamma Z) = {{\alpha\,G_F^2 m_{W}^2 m_{H}^3} \over {64\pi^2}} \bigg(1- {{m_Z^2} \over {m_H^2}} \bigg)^3 \bigg| \sum_{f} N_f {{Q_f \hat{v}_f}\over{c_W}} A_{1/2}^H(\tau_f,\lambda_f)+A_{1}^H(\tau_W,\lambda_W) \bigg|^2 \label{eq:part-width-hgaZ-sm}
\end{equation}
with $\alpha$ and $G_F$ are the fine-structure and the Fermi coupling constants respectively, $Q_f$ is the fermion electric charge defined in terms of the proton electric one, $c_W$ is the $cosine$ of the electroweak angle ($c_W^2=1-s_W^2$) and $\hat{v}_f$ is a dimensionless quantity related to the weak isospin of the fermion as $\hat{v}_f=2 I_f^3-4 Q_f s_W^2$. The two form factors are parameterized in \cite{Gunion:1989we} as follows:

\begin{align}
A_{1/2}^H(\tau_f,\lambda_f) & = \big[ I_1(\tau,\lambda) - I_2(\tau,\lambda) \big] , \\
A_{1}^H(\tau_W,\lambda_W) & =  c_W \bigg\{ 4 \Big( 3 - {s_W^2 \over c_W^2} \Big) I_2(\tau,\lambda) + \Big[ \Big( 1 + {2 \over \tau} \Big) {s_W^2 \over c_W^2} - \Big( 5 + {2 \over \tau} \Big) \Big]  I_1(\tau,\lambda) \bigg\} ,
\label{kappa}
\end{align}
where $\tau_i=4m_i^2/m_H^2$, $\lambda_i=4m_i^2/m_Z^2$ and the functions of two variables $I_1(\tau,\lambda)$ and $I_2(\tau,\lambda)$ are given in the Appendix \ref{appendixA}. These form factors  could also be expressed using the Passarino-Veltman scalar functions \cite{Passarino:1978jh}.

\begin{figure}[!h]
	\center 
	\includegraphics[width=0.6\textwidth]{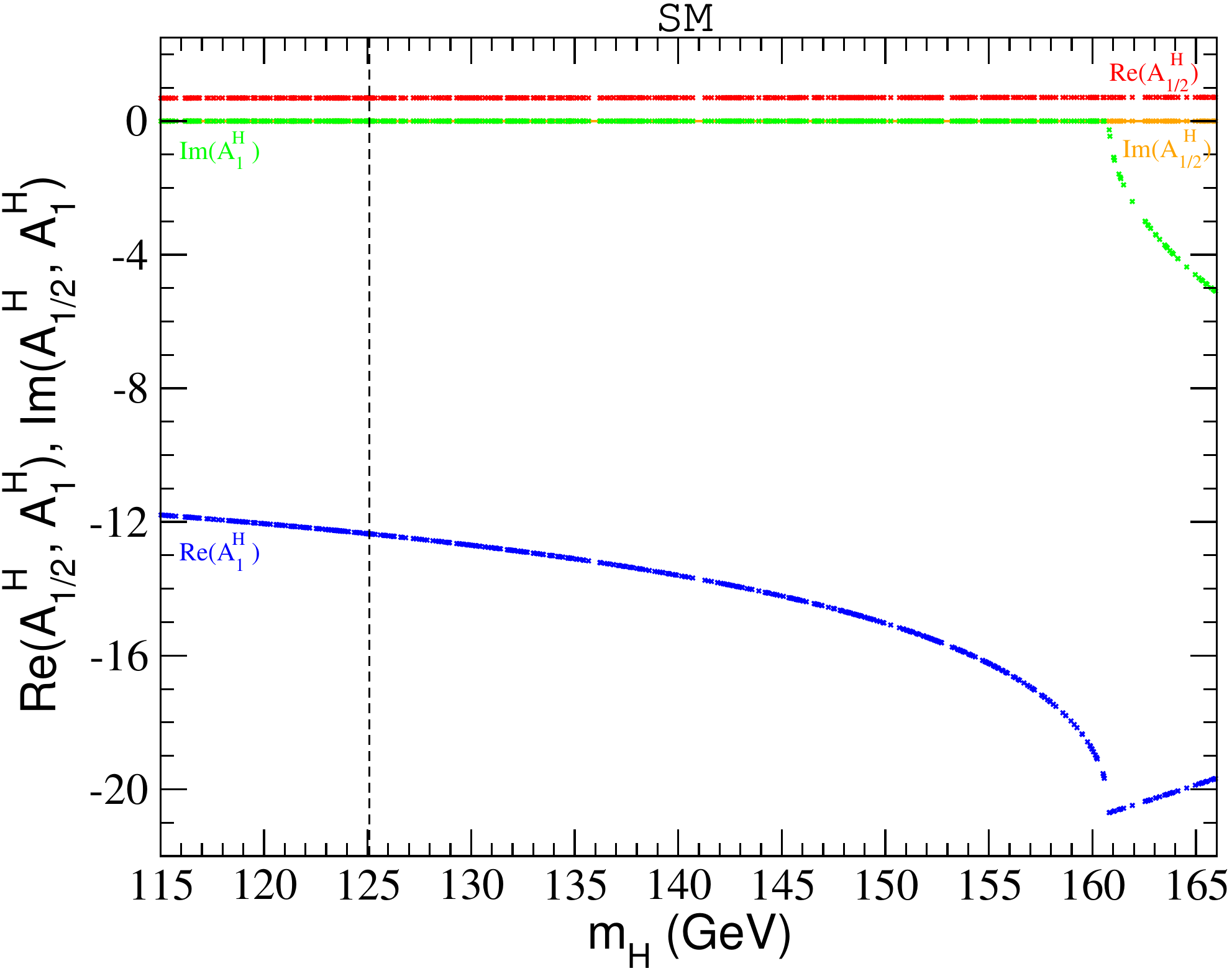}
	\caption{ Real and imaginary components  of the factors forms of $H \to  \gamma Z$ as a function of the SM Higgs Boson's  mass $m_H$.}
	\label{fig:contributions-sm}
\end{figure}

\noindent
For a better understanding of the various contributions between brackets in Eq.(\ref{eq:part-width-hgaZ-sm}), we illustrate graphically the real and imaginary  components of the form factors $A_{1/2}^H(\tau_f,\lambda_f)$ and $A_{1}^H(\tau_W,\lambda_W)$ as a function of the Higgs mass, based on the actual limit $\approx 125.09$ GeV. This behavior is exhibited in Fig.\ref{fig:contributions-sm}, where besides the opposite signs for both fermionic and bosonic contributions, one could note the substantial predominance of loops mediated by $W^\pm$ in such a way that any further interference between the two will have a negative sign, which implies a destructive interference. Moreover, for all mass values lying above the observed 125.09 GeV up to the $W^+W^-$ threshold, the partial width of the $H \to \gamma Z$ remains negligible compared to others, such as $H \to \gamma\gamma$ and $H \to b\bar{b}$. Last but not least, it may be mentioned that the observed Higgs boson won't share all the secrets of its creation mechanism at once, and hence, measuring and studying its proprieties remains a challenge for physicists. 
\begin{figure}[!ht]
	\center 
	\includegraphics[scale=1]{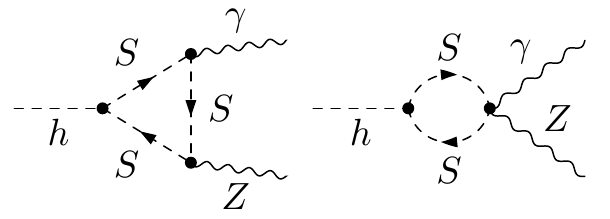}
	\caption{The new generic Feynman diagrams for $h \to  \gamma Z$ BSM, $S$ stands for charged particles.}
	\label{fig:diagrams-NP}
\end{figure}

\noindent
In this context, any theory BSM of particle physics may give rise to significant explanations of many experimental observations. Hence, we investigate in the next sections the modification of the previous analyses against the new physics contributions. This latter is highlighted through the new Feynman diagrams due to the existence of a newly charged scalar. The new diagrams are displayed in Figure \ref{fig:diagrams-NP}, where the loops are mediated, in addition to the modified previous contribution of the SM, by the charged ones.

\noindent
Concerning the decay width, the expression given previously by Eq.(\ref{eq:part-width-hgaZ-sm}), under these new contributions, takes the form :
\begin{align}
\Gamma(h \to \gamma Z) &= {{\alpha\,G_F^2 m_{W}^2 m_{h}^3} \over {64\pi^2}} \bigg(1- {{m_Z^2} \over {m_h^2}} \bigg)^3 \bigg| \sum_{f} \kappa_{f} N_f {{Q_f \hat{v}_f}\over{c_W}} A_{1/2}^h(\tau_f,\lambda_f)\nonumber\\
& + \kappa_W A_{1}^h(\tau_W,\lambda_W) + \sum_{S} Q_S^2 \frac{m_W^2}{m_{S}^2} \kappa_{ZSS} \kappa_{hSS} A_{0}^h(\tau_S,\lambda_S) \bigg|^2.
\label{eq:part-width-hgaZ-htm}
\end{align}
where, $S$ refers to charged scalar, $Q_S$ its electric charge, $\kappa_f$ and $\kappa_W$ are the reduced coupling of the Higgs boson $h$ to fermions and $W$ gauge boson respectively, while $\kappa_{ZSS}$ and $\kappa_{hSS}$ denotes the $Z$ and Higgs boson $h$ couplings to charged state. \textcolor{black}{The expression of $A_0^h(\tau_S,\lambda_{S})$ is given in literature\cite{Gunion:1989we} as follows:
\begin{align}
A_0^h(\tau_S,\lambda_S) = \left(\frac{1-2c_W^2}{c_W s_W}\right) I_1(\tau_{S},\lambda_{S})
\end{align}}
This decay width could be confronted with the BSM prediction via  the signal strength measurement, $\mu_{\gamma Z}$, defined by,

\begin{eqnarray}
\mu_{\gamma Z}^{NP} = \frac{\sigma(pp \to h \to 
	\gamma Z)^{\textrm{NP}}}{\sigma(pp \to h_{\textrm{SM}} \to \gamma Z)} = \frac{Br(h \to \gamma Z)^{\textrm{NP}}}{Br(h_{\textrm{SM}} 
	\to \gamma Z)} \times 
\left\{
\begin{array}{l}
1 \hspace{2.55cm}  (\text{Inert})\\
{{\Gamma(h \to gg)^{\textrm{NP}}}\over{\Gamma(h \to gg)^{\textrm{SM}}}} \hspace{1.0cm}  (\text{others})
\end{array}
\right.\label{eq:widthBSM}
\end{eqnarray}
Highlighting the areas where $\mu_{\gamma Z}$ and its most recent LHC measurement are consistent. \textcolor{black}{The latest experimental values (with uncertainties) of the signal strengths $\mu_{\gamma Z}$, $\mu_{\gamma \gamma}$, and $\mu_{ggh}$(production) at LHC, which have been used in the parameter scans are set in Table \ref{experimental_signal_strenghts}.
\begin{table}[H]
	\centering
\begin{tabular}{|c|c|c|c|}
\hline	
Signal strengths&$\mu_{\gamma Z }$&$\mu_{\gamma \gamma}$&$\mu_{ggh}$\\\hline
ATLAS&$2.0^{+1.0}_{-0.9}$\cite{Aad:2020plj}&$0.99_{-0.14}^{+0.15}$\cite{ATLAS:2018hxb}&$1.04_{-0.09}^{+0.09}$\cite{ATLAS:2019slw}\\\hline
CMS&$2.4_{+0.9}^{-0.9}$\cite{CMS:2021wog}&$1.18_{-014}^{+0.17}$\cite{CMS:2018piu}&$1.04_{-0.09}^{+0.09}$\cite{CMS:2020gsy}\\\hline
\end{tabular}
\caption{Experimental signal strengths $\mu_{\gamma Z }, \mu_{\gamma \gamma}$ and $\mu_{ggh}$(production) with uncertainties.}
\label{experimental_signal_strenghts}
\end{table}}
\section{Confronting $ZH\gamma$ BSM}
\label{sec:hgammaZ-bsm}
As outlined above, the Higgs decay into a $Z$ boson and a photon might be amended in extensions BSM either by the charged new particle contribution in the loop or by drastically changing couplings of the SM-like Higgs boson. To study such a purpose, three non-supersymmetric different models are taken into consideration, which are: the CP-conserving Two-Higgs doublet model (2HDMs), the Inert doublet model (IDM), and the Higgs Triplet model (HTM) with the hypercharge $Y=2$.  Except for the IDM, where the SM-like Higgs couplings are exact replicas of SM ones, all the remaining studied extensions modify the couplings to SM fermions and weak gauge bosons due to the mixing (the interaction) between the scalar fields. In what follows, in a nutshell, we go through the three models in order, putting the spotlight on the appropriate parameters for each model and achieving the land for a deep discussion on $\gamma V(V=\gamma, Z)$ excess explanation in the next section.
\subsection{Two-Higgs Doublet Model: \thdm s}
\label{sec:thdm-extension}
By extending the SM Higgs sector with a complex $SU(2)$ doublet $H_2$, the simplest extension of SM, known as 2HDM\cite{Aoki:2009ha,Branco:2011iw,Craig:2013hca} , is accomplished. Its most general gauge-invariant under $SU(2)_L\otimes U(1)_Y$, renormalizable scalar potential is provided by:
\begin{eqnarray}
V(H_1,H_2) &=&  m_{11}^2H_1^\dagger H_1+m_{22}^2H_2^\dagger H_2-[m_{12}^2H_1^\dagger H_2+{\rm h.c.}] + \frac{\lambda_1}{2}(H_1^\dagger H_1)^2
+\frac{\lambda_2}{2}(H_2^\dagger H_2)^2 \nonumber \\
&+& \lambda_3(H_1^\dagger H_1)(H_2^\dagger H_2)+\lambda_4(H_1^\dagger H_2)(H_2^\dagger H_1) +\frac{1}{2}[\lambda_5~(H_1^\dagger H_2)^2 +~{\rm h.c.}]
\label{eq:pot2HDM}
\end{eqnarray}
In which we suppose that the Flavour Changing Neutral Currents (FCNC) are suppressed by enforcing the $V(H_1,H_2)$ to be $\zt$ symmetric under the transformations: $H_1 \to -H_1$, $H_2 \to H_2$ on the potential, making the model a cp-conserving one. The two doublets can be parameterized as follows using Electroweak Symmetry Breaking (EWSB): 
\begin{eqnarray}
H_1 =
\begin{bmatrix}
\phi_1^+ \\
(v_1+\rho_1+i\eta_1)/\sqrt{2} 
\end{bmatrix}
\quad 
{\rm and}
\quad 
H_2 =
\begin{bmatrix}
\phi_2^+ \\
(v_2+\rho_2+i\eta_2)/\sqrt{2} 
\end{bmatrix},
\label{representa-htm}
\end{eqnarray}
where no CP violation is assumed in the vevs $v_1$ and $v_2$ and the EW scale is fixed, i.e.,  $v_{EW}=\sqrt{v_1^2+v_2^2}=(\sqrt{2}G_F)^{-{1\over2}}=246$ GeV.

The two doublets $H_{1,2}$ have $8$ real degrees of freedom, $3$ are gauged while the remained fields originate 5 Higgs bosons: three electrically neutral, two refer to scalar ($h$ and $H$) and the third one is pseudoscalar ($A$); the remaining two bosons are electrically charged ($H^\pm$). The latter is the new charged particles that might contribute to the deviations of the $\gamma V (V= \gamma, Z)$ channel events. Both states $h$ and $H$ could mimic the observed 125.09 GeV, but in what follows in our study, we will concentrate on the scenario where the lightest $h$ is the SM-like, and we refer to this as the $h-$ scenario, respectively. Therefore, in order to completely discuss the scalar sector of 2HDMs, we conventionally use the subsequent set of parameters:
\begin{equation}
\mathcal{P}_\text{2HDMs}=\left\{m_h=125.09\ \rm{GeV},\,m_H,\,m_A,\,m_{H^\pm},\,m_{12}^2,\,\tan\beta,\,\sin(\beta-\alpha)\right\}
\label{eq:inputpara-2hdm} 
\end{equation}
Four distinct Higgs-fermion interaction configurations result from the various quark field transformations under the $\zt$ symmetry: Only one doublet (supposed here $H_1$) couples to all fermions in the first version of the model (denoted 2HDMs Type-I) and generate masses to all fermions; in the second version or Type-II, one doublet (denoted $H_2$) couples to the up quark while the other $H_1$ doublet couple to the down quark and leptons, generating the masses to fermions; the third version denoted Type-X or Lepton Specific couples the doublet $H_2$ to all quarks while the other doublet $H_1$ couples to the charged leptons; the last version of interaction carried the name Type-Y (or Flipped) couples the doublet $H_2$  to up-type quarks and leptons while the other doublet $H_1$ couples to down-type quarks. The Higgs and fermion sectors interact according to the Yukawa Lagrangian, which is expressed in the mass eigenstate basis by,
\begin{eqnarray}
- {\mathcal{L}}_{\rm Yukawa} = \sum_{f=u,d,l} \left(\frac{m_f}{v} \kappa_f^h \bar{f} f h + 
\frac{m_f}{v}\kappa_f^H \bar{f} f H 
- i \frac{m_f}{v} \kappa_f^A \bar{f} \gamma_5 f A \right) + \nonumber \\
\left(\frac{V_{ud}}{\sqrt{2} v} \bar{u} (m_u \kappa_u^A P_L +
m_d \kappa_d^A P_R) d H^+ + \frac{ m_l \kappa_l^A}{\sqrt{2} v} \bar{\nu}_L l_R H^+ + H.c. \right),
\label{Yukawa-1}
\end{eqnarray}
where all the $\kappa_{f}^{h_i}$ factors are grouped in table \ref{couplage}.

{\renewcommand{\arraystretch}{1.5} 
{\setlength{\tabcolsep}{0.2cm} 
	\begin{table}[H]
		\centering
		\begin{tabular}{|c|c|c|c|c|c|c|c|c|c|}
			\hline
			Type&$\kappa_{u}^{h}$&$\kappa_{d}^{h}$&$\kappa_{l}^{h}$&$\kappa_{u}^{H}$&$\kappa_{d}^{H}$&$\kappa_{l}^{H}$&$\kappa_{u}^{A}$&$\kappa_{d}^{A}$&$\kappa_{l}^{A}$\\\hline
			I&$\frac{c_\alpha}{s_\beta}$&$\frac{c_\alpha}{s_\beta}$&$\frac{c_\alpha}{s_\beta}$&$\frac{s_\alpha}{s_\beta}$&$\frac{s_\alpha}{s_\beta}$&$\frac{s_\alpha}{s_\beta}$&$\cot\beta$&$-\cot\beta$&$-\cot\beta$\\\hline
			II&$\frac{c_\alpha}{s_\beta}$&$-\frac{s_\alpha}{c_\beta}$&$-\frac{s_\alpha}{c_\beta}$&$\frac{s_\alpha}{s_\beta}$&$\frac{c_\alpha}{c_\beta}$&$\frac{c_\alpha}{c_\beta}$&$\cot\beta$&$\tan\beta$&$\tan\beta$\\\hline
			X&$\frac{c_\alpha}{s_\beta}$&$\frac{c_\alpha}{s_\beta}$&$-\frac{s_\alpha}{c_\beta}$&$\frac{s_\alpha}{s_\beta}$&$\frac{s_\alpha}{s_\beta}$&$\frac{c_\alpha}{c_\beta}$&$\cot\beta$&$-\cot\beta$&$\tan\beta$\\\hline
			Y&$\frac{c_\alpha}{s_\beta}$&$-\frac{s_\alpha}{c_\beta}$&$\frac{c_\alpha}{s_\beta}$&$\frac{s_\alpha}{s_\beta}$&$\frac{c_\alpha}{c_\beta}$&$\frac{s_\alpha}{s_\beta}$&$\cot\beta$&$\tan\beta$&$-\cot\beta$\\\hline
			\end{tabular}
			\caption{The Yukawa coupling coefficients of the neutral Higgs bosons $h$, $H$, and $A$ to quarks ($u$, $d$) and charged leptons ($l$) in the four 2HDMs types.}
			\label{couplage}
		\end{table}
The Higgs-fermion couplings in the fermion mass basis are flavor diagonal and completely dependent on the mixing angle $\beta$ in the alignment limit in the absence of FCNCs, which is also an important point to keep in mind. In such a case, the coefficients $\kappa^{h}_f$ are redefined using the following trigonometric simplification expressions,
\begin{align}
\kappa_{u,d}^{I,X} = \kappa_{u}^{II,Y} &= \frac{\cos\alpha}{\sin\beta} = \sin(\beta-\alpha)+\cos(\beta-\alpha)\cot\beta\label{coupling-h1}\\
\kappa_{d}^{II,Y} &= -\frac{\sin\alpha}{\cos\beta} = \sin(\beta-\alpha)-\cos(\beta-\alpha)\tan\beta\label{coupling-h2}
\end{align}
As far as the vectorial bosons, $W$ and $Z$, their reduced couplings to the Higgs scalars are given by,
\begin{equation}
\kappa_{hVV} = \sin(\beta-\alpha),\qquad
\kappa_{HVV} = \cos(\beta-\alpha),\qquad 
\kappa_{AVV} = 0
\label{eq:gaugecouplings} 
\end{equation}
where $VV=W^+W^-\,\text{or}\,ZZ^{\ast}$. 
\noindent
In the $h-$scenario, where we approach the alignment limit, the normalized coupling should go near the unity, so $\cos(\beta-\alpha) \rightarrow 0$. however, for a large $\tan\beta$, we read from eq.\ref{coupling-h2} that the such limit is only reachable for a very small $|\cos(\beta-\alpha)\tan\beta| \ll 1$ in the Type-II and Type-Y. Concerning the couplings of the charged scalars, $H^{\pm}$, to $h$, it is given by,
\begin{eqnarray}
\kappa_{hH^+H^-} = \frac{1}{v}\times\left[(m_h^2-2 m_{H^\pm}^2)s_{\beta-\alpha}-\frac{2c_{\beta+\alpha}}{s_{2\beta}^2}(m_h^2 s_{2\beta}-2 m_{12}^2)\right] \label{eq:coupSHpHm1}
\end{eqnarray}
and, although, this charged state is not yet discovered, its effects might be detected indirectly through loop contributions involving couplings given by Eqs.(\ref{eq:coupSHpHm1}), such as $h \to \gamma\gamma$ and $h \to \gamma Z$ that already existed in the SM. Due to the presence of a light particle in the loop, this is feasible if $m_{H^\pm} \sim v$.
When the $H^\pm$ loop contribution gets close to a constant (i.e. $m_{H^\pm} >> v$), this is also conceivable\cite{Arhrib:2003vip,Das:2015gyv}. But increasing $m_{H^\pm}$ too much will call for quartic couplings that are outside the unitarity bounds. Nevertheless, there is still a sizable range of $H^\pm$ masses where the charged Higgs contributions to $h\to \gamma\gamma$ and $h\to \gamma Z$ may be found. 

The model has seven free, real, independent parameters, as we have shown in Eq. \ref{eq:inputpara-2hdm}. However, the theory is not particularly predictive with seven unrestricted factors. It is then crucial to limit the parameter space using both the existing experimental data and theoretical constraints, as is the case with any extension of BSM. On the theoretical side, we have implemented the positivity, perturbativity, and unitarity constraints, while on the experimental side, we have enforced the EWPOS through the parameter $S$, $T$, and $U$ , collider bounds, Higgs signal strength, and flavor physics constraints as follows:
\begin{itemize}
	\item {\bf Positivity constraints}, which assure that the potential is constrained from below \cite{Barroso:2013awa,Deshpande:1977rw} when the scalar fields get extremely large, this corresponds to,
	\[\lambda_1 > 0\,\quad \lambda_2 > 0\,\quad \lambda_3 > -\sqrt{\lambda_1\lambda_2}\,\quad \lambda_3+\lambda_4-|\lambda_5| > -\sqrt{\lambda_1\lambda_2}. \]
	\item {\bf Perturbative unitarity constraints} : they put bounds on the amplitude of partial waves, which in turn curb the values of the coupling constants that appear in the $\mathcal{S}-$matrix eigenvalues, $x_i$. At high energy Higgs-Higgs scattering, those $|x_i|<8\pi$, and could be expressed at tree level as in \cite{Akeroyd:2000wc,Kanemura:1993hm,Arhrib:2000is}.
	\item {\bf Electroweak precision constraints}, or the so-called oblique parameters $S$, $T$, and $U$. They're related to the new physics in electroweak radiative corrections. In the 2HDM, a common set of these parameters is given in \cite{Branco:2011iw,Grimus:2007if,Grimus:2008nb}, and the corresponding expressions are used throughout our study. 
\end{itemize}
To comply fully therewith, we have used the public tool 2HDMC-1.8~\cite{Eriksson:2009ws} to check the aforementioned theoretical constraints, and we have resorted to certain public codes to check the agreement with the constraints stemming from the existing experimental searches, such as HiggsBounds-5.10.2 \cite{Bechtle:2020pkv,Bechtle:2013wla} which takes into consideration the exclusion bounds from Higgs searches at LEP, Tevatron and LHC experiments, HiggsSignal-2.6.2 \cite{Bechtle:2020uwn,Bechtle:2014ewa} to satisfy the agreement with the signal strength measurements of the observed Higgs at LHC, and SuperIso-v4.1 \cite{Mahmoudi:2008tp} to check the constraints from B-physics, the following is a list of relevant B-physics observables used in this work.
\begin{itemize}
	\item BR($B\to X_{s}\gamma)_{E_\gamma\ge 1.6\ \text{GeV}}$ = $(3.32\pm 0.15)\times10^{-4}$,\cite{Haller:2018nnx,HFLAV:2016hnz}
	\item BR($B_{s}\to \mu^{+}\mu^{-}$) = $(3.0 \pm {0.6_{\text{stat}}}_{-0.2_{\text{sys}}}^{+0.3}) \times 10^{-9}$\cite{Haller:2018nnx,Arhrib:2018ewj},
	\item BR($B_{d}\to \mu^{+}\mu^{-}$) = $({1.5_{-1.0_{\text{stat}}}^{+1.2}}_{-0.1_{\text{sys}}}^{+0.2})\times10^{-10}$\cite{Haller:2018nnx,Mahmoudi:2008tp},
	\item BR($B_{\mu}\to \tau\nu$) = $(1.06 \pm 0.19) \times 10^{-4}$\cite{Haller:2018nnx}.
\end{itemize}
\subsection{Inert Doublet Model: \idm}
\label{sec:idm-extension}

Firstly suggested by Ma, the inert doublet model \cite{Deshpande:1977rw,LopezHonorez:2006gr,Barbieri:2006dq} is one of the simplest models where an inert scalar doublet $H_2$ extends the SM scalar sector, giving rise to the prospect of a stable dark matter. An exact $\zt$ symmetry is established in this framework, allowing all SM particles to be even while the double $H_2$ is odd. The only way the two doublets interact at the tree level is by mixing.  The IDM fields are parametrized as follows:
\begin{eqnarray}
H_1 = \left (\begin{array}{c}
G^\pm \\
\frac{1}{\sqrt{2}}(v + h + i G^0) \\
\end{array} \right)
\qquad , \qquad
H_2 = \left( \begin{array}{c}
H^\pm\\ 
\frac{1}{\sqrt{2}}(H^0 + i A^0) \\ 
\end{array} \right) 
\end{eqnarray}
Where $G^0$ and $G^\pm$ are the Nambu-Goldstone bosons consumed by the
longitudinal component of the weak gauge boson ($W^\pm$ and, $Z^0$).  $v$ is the SM Higgs doublet $H_1$'s vacuum expectation value, or VEV for short.  The $H_2$- fermions interaction is only present by mixing with $H_1$ in the IDM since the scalar doublet $H_2$ does not couple with the SM fermions. The potential that is the most universally renormalizable, gauge-invariant, and CP-invariant is provided by: 
\begin{eqnarray}
V(H_1,H_2) &=& \mu_1^2 |H_1|^2 + \mu_2^2 |H_2|^2  + \lambda_1 |H_1|^4
+ \lambda_2 |H_2|^4 +  \lambda_3 |H_1|^2 |H_2|^2\nonumber \\
&+& \lambda_4 |H_1^\dagger H_2|^2 + \frac{\lambda_5}{2} \left\{ (H_1^\dagger H_2)^2 + {\rm h.c} \right\} 
\label{eq:pot-idm} 
\end{eqnarray}
whereas, due to the potential's hermicity, all four parameters  $\lambda_i, i = 1, \cdots, 4$ are real. The fields $H_1$ and $H_2$ can be appropriately redefined to absorb the phase of $\lambda_5$, creating CP conservation in the scalar sector.
The spectrum of this potential will contain five scalar particles after the spontaneous symmetry breaking (SSB) of $SU(2)_L\otimes U(1)_Y$ down to $U(1)_Y$: three neutral scalar $h$, $H$ and $A$, where $h$  might be recognized as the SM Higgs boson discovered at LHC with a mass close to 125.09 GeV, along with two charged
scalars $H^\pm$ , are present. Their masses are:
\begin{equation}
\label{spect.IHDM}
m_h^2 = - 2 \mu_1^2 = 2 \lambda_1 v^2,\quad 
m_{H}^2 = \mu_2^2 + \lambda_L v^2,\quad 
m_{A}^2 = \mu_2^2 + \lambda_S v^2,\quad
m_{H^{\pm}}^2 = \mu_2^2 + \frac{1}{2} \lambda_3 v^2
\end{equation}
where the definition of $\lambda_{L,S}$ are: 
\begin{align}
\lambda_{L,S} = \frac{1}{2}\left(\lambda_3 + \lambda_4 \pm \lambda_5\right) 
\end{align}
Eight independent parameters, including five $\lambda$, two $\mu_i$ and $v$ make up this model. By virtue of the minimization condition, one parameter is removed, and the $W$ boson mass fixes the $VEV$.  Finally, six independent real parameters are all that is left to fully assess the model, chosen as follows :
\begin{eqnarray}
\left\{ \mu_2^2, \lambda_2, m_h=125.09\,\rm{\gev}, m_{H}, m_{A}, m_{H^\pm} \right\}
\label{param.IHDM}
\end{eqnarray}

\noindent
The trilinear Higgs couplings required for our analysis are listed below for completeness: 
\begin{equation}
\label{3couplings.IHDM}
\kappa_{hHH} = -2\,v\,\lambda_L,\quad \kappa_{hAA} = -2\,v\, \lambda_S,\quad \kappa_{hH^{\pm}H^{\pm}} = -v\,\lambda_3
\end{equation}

\noindent
Similar to the \thdm s, the \idm parameters should satisfy various theoretical and experimental constraints cited above in the \thdm s \ model (Sec. \ref{sec:thdm-extension}). Additionally, it's subject to the following theoretical and experimental criteria,
\begin{itemize}
	\item {\bf Existence of the inert vacuum} \cite{Deshpande:1977rw}, which is pertinent since the CP-conserving minimum could be a global one, and the following inequalities 
	\begin{equation}
	\label{vacuum.IHDM.2}
	m_h^2,\,m_H^2,\,m_A^2,\,m_{H^\pm}^2>0\,\quad \text{and}\quad \mu_1^2/\sqrt{\lambda_1}<\mu_2^2/\sqrt{\lambda_2}
	\end{equation}
	must be met thereon.  
\end{itemize}

\noindent
Experimentally, although there has been no hint of a charged particle in colliders yet, such a Higgs boson will assist in any way to anchor the allowed parameter space of the IDM. Accordingly, some branching rates are strongly influenced by its mass, as in the midst of the gauge bosons. To make sure that we exclude the decay channels, $W^\pm \rightarrow H H^{\pm}\ , A H^{\pm}$ and $Z \rightarrow H A,\  H^{\pm} H^{\mp}$, we set,
\begin{align}
m_{H, A}+m_{H^\pm} > m_{W^\pm},\ m_{H}+m_{A} > m_{Z},\ 2m_{H^\pm} > m_{Z}
\end{align}
and aside from that, the measurement of its upper limit lifetime, relatively sizeable $\tau\leq 10^{-7}s$, leads to a lower limit on the width of $\Gamma_{tot}(H^{\pm}) \ge 6.58\times 10^{-18}\,$ \gev. We also employ one significant revision of a LEP study\cite{EspiritoSanto:2003by} within the IDM framework \cite{Lundstrom:2008ai}. This in particular, eliminates any areas where, 
\begin{align}
m_A < 100\ \rm{\gev},\ m_H < 80\ \rm{\gev},\ \Delta  m(A,H) > 8\ \rm{\gev}.
\end{align}
\noindent
Motivated by dark matter (DM) explanations, the IDM scalar sector is, moreover, a good haven for such research, which has been strongly supported by many astrophysical and cosmological observations. Here we require agreement with the dark matter relic density obtained from the Planck experiment. \cite{Ade:2015xua}
\begin{align}
\Omega_{\rm DM}h^2 < \Omega_{c}h^2 = 0.1197 \pm 0.0022\label{plank_limit}
\end{align}
We also double-checked that the IDM parameter space matches the DM Direct Detection (DD) experiment boundaries. We used the micromegas v5.3.35\cite{Belanger:2020gnr} package to calculate the spin-independent cross-section of DM proton scattering, $\sigma_{SI}$. We employ the re-scaled direct detection cross section, $\hat{\sigma_{SI}}=R_{\Omega}\times \sigma_{SI}$, to illustrate the IDM results, where the scaling factor, $R_\Omega = \Omega_{\rm DM}/\Omega_{\rm DM}^{\rm{plank}}$, taking into consideration the scenario of $H$ representing only a portion of the entire DM budget \textcolor{black}{and not 100\% of the dark matter relic density}, allowing for an easy comparison of the model prediction with the XENON1T limitations\cite{XENON:2018voc}.
\subsection{Extension with real triplet: \htm}
\label{sec:htm-extension}
Instead of another doublet, as has been seen previously, an additional $SU(2)_{L}$ real triplet, $\Delta$ \cite{Magg:1980ut,Cheng:1980qt,Perez:2008ha,Arhrib:2011uy,Arhrib:2014nya,Agrawal:2018pci,Arhrib:2019kqb} with hypercharge $Y_{\Delta}=2$ is introduced to the SM Higgs sector. The Higgs potential is given by 
\begin{eqnarray*}
	V(H, \Delta) &=& -m_H^2{H^\dagger{H}}+\frac{\lambda}{4}(H^\dagger{H})^2+M_\Delta^2\tr(\Delta^{\dagger}{\Delta})
	+[\mu(H^T{i}\sigma^2\Delta^{\dagger}H)+{\rm h.c.}]\nonumber\\
	&+& \lambda_1(H^\dagger{H})\tr(\Delta^{\dagger}{\Delta})+\lambda_2(\tr\Delta^{\dagger}{\Delta})^2
	+\lambda_3\tr(\Delta^{\dagger}{\Delta})^2 +\lambda_4{H^\dagger\Delta\Delta^{\dagger}H}
	\label{eq:Vpot-htm}
\end{eqnarray*}
in which the two Higgs multiplets are parameterized by,  
\begin{eqnarray}
\Delta &=\begin{bmatrix}
{\delta^+ \over \sqrt{2}} &  \delta^{++} \\
\delta^0 & -{\delta^+ \over \sqrt{2}}
\end{bmatrix}
\quad 
{\rm and}
\quad 
H=
\begin{bmatrix}
\phi^+ \\
\phi^0 
\end{bmatrix}.
\label{representa-htm}
\end{eqnarray}
Also, it should be noted that such model is characterized by a lepton number that is not zero, $L=2$, violated explicitly by the $\mu$ parameter
and related to the presence of the doubly-charged state $\delta^{\pm\pm}$. 
After the electroweak symmetry breaking (EWSB), the components $\phi^0$ and $\delta^0$ get a non-vanishing vacuum expectation value ({\it vev}'s), $\langle\phi^0\rangle \to v_H$ and $\langle\delta^0\rangle \to v_\Delta$, yielding two vacuum conditions \cite{Arhrib:2019kqb}
and so, seven mass eigenstates appear, namely $H^{\pm\pm}$, $H^\pm$, $A$, $H$ and $h$. Its mass spectrum can be represented in terms of the Lagrangian's parameters as, 
\begin{eqnarray}
&& m_{H^{\pm\pm}}^2  =  \frac{\sqrt{2}\mu{v_H^2}-\lambda_4v_H^2v_\Delta-2\lambda_3v_\Delta^3}{2v_t},\label{eq:mHpmpm}\\
&& m_{H^{\pm}}^2 =  \frac{(v_H^2+2v_\Delta^2)\,[2\sqrt{2}\mu-\lambda_4v_\Delta]}{4v_\Delta},\label{eq:mHpm}\\
&& m_{A}^2 =  \frac{\mu(v_H^2+4v_\Delta^2)}{\sqrt{2}v_\Delta},\label{eq:mA0} \\
&& m_H^2 = \frac{1}{2}\Big(\lambda v_H^2 s_\alpha^2 + c_\alpha^2 \Big[\sqrt{2}\mu\frac{v_H^2}{v_\Delta}\big(1+4\frac{v_\Delta}{v_H}\tan\alpha\big) +4v_\Delta^2\big(\lambda_{23}^+ - \lambda_{14}^+\frac{v_H}{v_\Delta}\tan\alpha\big) \Big]\Big),\\
&& m_h^2 = \frac{1}{2}\Big(\lambda v_H^2 c_\alpha^2 + s_\alpha^2 \Big[ \sqrt{2}\mu\frac{v_H^2}{v_\Delta}\big(1-4\frac{v_\Delta}{v_H\tan\alpha}\big) +4v_\Delta^2\big(\lambda_{14}^+\frac{v_H}{v_\Delta\tan\alpha}+\lambda_{23}^+\big)\Big]\Big),
\end{eqnarray}
where $\alpha$ represents the sector's mixing angle in the CP-even case\cite{Arhrib:2011uy}. Therefore, in our numerical evaluation, we deal with the case where the lightest Higgs boson $h$ represents the SM-like, which sets the Higgs self coupling at a value of $\lambda \sim 0.52$. We also assume $v_\Delta = 1$ \gev\ and $\mu$ around $0.8 \sim 1.2$ \gev.

\noindent
Like in the previous extensions, the parameter space of \htm is being controlled under the guidance of both theoretical and experimental constraints. Such requirements have been entirely examined in the literature, and we summarize their results. These are:
\begin{itemize}
	\item {\bf Vacuum constraints}: obviously, the minimization equations \cite{Arhrib:2011uy} warrant that the vacuum is extremely local, without specifying if the hypothetical electroweak vacuum is a minimum, a saddle point, or a local maximum. Accordingly, the non-linearity of those equations reflects the existence of different values $(\langle\phi^0\rangle,\,\langle\delta^0\rangle )$ in the allowed space, and for only one pair $(v_H,\,v_\Delta)$, the $V(H,\Delta)$ is in a minimum situation. Such scrutiny can be very involved with regard to the ten independent parameters in the scalar potential. As a result, assuming that the EWSB would be energetically favored if $\langle V(H,\Delta) \rangle < 0$, the condition,
	\begin{equation}
	4 \sqrt{2} \mu v_\Delta - \lambda v_H^2 - 4 (\lambda_1+\lambda_4) v_\Delta^2 < 0
	\label{eq:vacuum-htm}
	\end{equation}
	is required and sufficient enough for the minimum to be unique \cite{Arhrib:2011uy}.
	\item {\bf Positivity constraints}: the following requirements must be fulfilled in order to keep the scalar potential $V(H, \Delta)$ constrained from below,  \cite{Arhrib:2011uy,Bonilla:2015eha}:
	\begin{eqnarray}
	&& \lambda \geq 0 \;\;{\rm \&}  \;\; \lambda_{23}^+ \geq 0  \;\;{\rm \&}  \;\;\tilde\lambda_{23}^+ \geq 0  \;\;{\rm \&} \;\;\lambda_1+ \sqrt{\lambda \lambda_{23}^+} \geq 0 \;\;{\rm \&}\;\;\lambda_{14}^++ \sqrt{\lambda \lambda_{23}^+} \geq 0 \label{eq:bound1} \nonumber\\
	&& {\rm \&} \;\; 2 \tilde{\lambda}_{14}+\sqrt{(2\lambda \lambda_3-\lambda_4^2) (2\frac{\lambda_2}{\lambda_3} + 1)} \geq 0 \;\; {\rm or} \;\; \lambda_3 \sqrt{\lambda} \le |\lambda_4| \sqrt{\lambda_{23}^+} \label{eq:bound2}
	\end{eqnarray}
	\item {\bf Perturbative unitarity requirement}: it is related to the scattering matrix that relates the initial and final states of a physical system undergoing a 2-by-2 scattering process. Such an S-matrix is defined as the unitary, which constrains the amplitude of partial waves \cite{Arhrib:2011uy}, which therefore places restrictions on the coupling constants' values. Thus, at high energy, all the eigenvalues $\Lambda_{i}$ of the scattering matrix must be $<8\pi$ in absolute value.
	\item {\bf Electroweak Precision Observables}: Firstly introduced by Peskin and Takeuchi \cite{Peskin:1990zt}, the oblique parameter's role is evident in the parametrization of new physics contribution to electroweak radiative corrections, including the total $Z$ coupling strength which is related to $\Delta \rho \propto T$. That is, in the framework of \htm, the major contribution to the $T$-parameter derives from loops that involve $H$, $A$, $H^\pm$ or $H^{\pm\pm}$. In our scenario where $m_{h}\approx\,125$ GeV ($c_\alpha \approx 1$), the new triplet-like scalars contribution reads \cite{Das:2016bir}
	\begin{align}
	T_{HTM} &=  \frac{1}{4\pi s_w^2m_W^2} \Big[ F(m_{H^\pm}^2,m_{A}^2) + F(m_{H^{\pm\pm}}^2,m_{H^{\pm}}^2)\Big],
	\label{eq:Tpara-htm}
	\end{align}
	where $F\left(I,J \right)$ is a mathematical function given by \cite{Grimus:2008nb}
	\begin{align}
	F\left( I, J \right) &= 
	\left\{ \begin{array}{lcl}
	\displaystyle{
		\frac{I + J}{2} - \frac{I J}{I - J}\, \ln{\frac{I}{J}}
	}
	& \Leftarrow & I \neq J,\\*[1mm]
	0 & \Leftarrow & I = J.
	\end{array} \right.
	\end{align}
	\noindent
	Experimentally, the global fit yields \cite{Tanabashi:2018oca}
	\begin{equation}
	T=0.09 \pm 0.07
	\label{eq:Tpara }
	\end{equation}
	and the HTM fully belongs within the $2\sigma$ allowed region of this oblique parameter.
	\item {\bf Collider experiments constraints}: 
	The current direct limits on the mass of the charged and doubly-charged Higgs boson in the BSM are set by the LEP and Tevatron, while the Large Hadron Collider (LHC) sets indirect limits that depend on the decay channel and the specific type of model considered. The LEP experiment excludes the mass of charged Higgs below $80~\rm{GeV}$ at 95$\%$ CL in 2HDM type-II, considering only the decays $H^+ \to c\bar{s}$ and $H^+ \to \tau \nu_{\tau}$ with BR$(H^+ \to c\bar{s}) + \mathrm{BR}(H^+ \to \tau \nu_{\tau}) = 1$ \cite{LEPHiggsWorkingGroupforHiggsbosonsearches:2001ogs}. Such a bound gets stronger and reach $92$ GeV (at $95\%$ CL), if BR$(H^+ \to \tau \nu_{\tau}) = 1$. Search for the
decay mode $H^+ \to AW^+$ with $A \to b\bar{b}$, which is not negligible in 2HDM type-I, leads to the corresponding $m_{H^{+}}$ limit of $72.5$ GeV ($95\%$ CL) if $m_A > 12$ GeV \cite{Logan:2009uf}. For a charged Higgs mass of $100$ GeV, searches at the Tevatron based on top anti-top pair production with the subsequent $t \to bH^+$ decay (setting BR$(H^+ \to \tau^+ \nu) = 1$) have set a limit on BR$(t \to bH^+)$ to be less than 0.2 \cite{D0:2009hbc}. LHC searches for $H^\pm$ have set an upper limit at $95\%$ confidence level on the production cross section multiplied by the branching ratio, $\sigma(pp \rightarrow H^\pm tb)\times \rm{BR}(H^\pm \rightarrow tb)$, which ranges from 3.6 (2.6) pb at $M_{H^\pm} = 200$ GeV to $0.036~(0.019)$ pb at $M_{H^\pm} = 2$ TeV. 
\end{itemize}
In addition to the direct searches, collider searches for neutral Higgs bosons can also result in indirect, model-dependent limits on the charged Higgs boson.
The $95\%$ C.L. limits from all experimental searches mentioned above are included in
our studies with HiggsBounds5.10.2. These limits exclude a charged Higgs boson with mass below $86.57$ GeV in 2HDMs type-I and $95.13$ GeV in 2HDMs type-X. In type-II and Y, we have taken the theoretical limit on the charged Higgs boson mass by default in our scans, which is $580$ GeV. (See appendix for more details.)
By embedding charged states, notably the $H^{\pm\pm}$, in its scalar spectrum, the \htm model could be the most recognizable groundwork for searching new physics. Meanwhile, $v_\Delta$ might be a decisive point and depending on its values, spectacular signatures could be envisaged. Indeed, while the gauge boson mode $H^{\pm\pm} \to W^\pm W^\pm $ takes over for larger triplet $vev$($v_{\Delta}$), the $H^\pm$ mostly decays into the same-sign leptonic states $H^{\pm\pm} \to l^{\pm}l^{\pm}$ for smaller triplet $vev$\cite{Perez:2008ha,Melfo:2011nx}. Numerous investigations have been suggested to investigate the possibility of doubly charged Higgs production and decay either at the LEP \cite{Abdallah:2002qj} or at the LHC \cite{Perez:2008ha,Melfo:2011nx,delAguila:2008cj,Mitra:2016wpr,Agrawal:2018pci}. For a small triplet vev $v_t < 10^{-4}$ GeV, the most recent experimental lower limits coming from pair-production,  reported by ATLAS is $m_{H^{\pm \pm}} \geq 870$ GeV at 95$\%$ C.L. \cite{ATLAS:2017xqs}, while the associated production $pp \to H^{\pm\pm}H^\mp$ and the subsequent decay, $H^{\pm}\to l^\pm \nu$ performed by CMS gives the stringent constraint $m_{H^{\pm \pm}} \geq 820$ GeV at 95$\%$ C.L. for $e,\mu$ flavor\cite{CMS:2017pet}. However, one can find scenarios where the mass goes down to $90$ - $100$ GeV for small $v_t$. In addition, L3, OPAL and Delphi LEP-II experiments assume that $H^{\pm\pm}$ decay dominantly to a pair of leptons. This forces the doubly charged Higgs mass to be $M_{H^{\pm\pm}}>97.3$ GeV \cite{DELPHI:2002bkf} at 95$\%$ C.L.

In the \htm, the reduced coupling of the Higgs boson $h$ to fermions $\kappa_{f}$, $W$ gauge bosons  $\kappa_{W}$ as well as its couplings together with the $Z$-gauge boson to either single- or doubly- charged Higgs $\kappa_{hSS}$, $\kappa_{ZSS}$ read respectively :
\begin{align}
&\kappa_{f} = \frac{c_\alpha}{c_{\beta^\pm}},\qquad\qquad\qquad  
\kappa_{W} = c_\alpha c_{\beta^\pm} + \sqrt{2} s_\alpha s_{\beta^\pm},\qquad\qquad\quad 
\kappa_{hH^{\pm}H^{\mp}}  \simeq -\frac{1}{2}\big(2 \lambda_1 +\lambda_4 \big) v_H, \nonumber\\
&\kappa_{hH^{\pm\pm}H^{\mp\mp}}  \simeq - \lambda_1 v_H,\quad  
\kappa_{ZH^{\pm}H^{\mp}} = \frac{s_{\beta^\pm}^2 c_{\theta_w}^2 - (1+c_{\beta^\pm}^2)s_{\theta_w}^2}{c_{\theta_w} s_{\theta_w}},\quad  
\kappa_{ZH^{\pm\pm}H^{\mp\mp}}  = \frac{1- 2 s_w^2}{c_w s_w}. \nonumber
\label{eq:coupling-htm}
\end{align}
where $\theta_w$ and $\beta^\pm$ are respectively Weinberg and the charged sector mixing angles \cite{Arhrib:2011uy}.
\section{Numerical Results }
\label{sec:res}
In this section, we describe the various observables subject to our calculations and show the numerical results in the three previously discussed models.
\subsection{In the \thdm s}
\label{sec:thdm-res}  
We study the $h \to \gamma\,V$ ($V=\gamma, Z$) decays in the cp-conserving \thdm, and we compare these decays with the ones of the SM to understand what changes such extensions imply. So, drawing upon the $h-$ scenario, the parameter spaces are set in Table. \ref{scanh}.
{\renewcommand{\arraystretch}{1.1} 
{\setlength{\tabcolsep}{0.04cm}
\begin{table}[H]
\centering
\begin{tabular}{|l|c|c|c|c|c|c|c|c|}
	\hline
		Parameters&$m_h$&$m_H$&$m_A$&$m_{H^\pm}$&$\tan\beta$&$\sin(\beta-\alpha)$&$m_{12}^2$(GeV$^2$)&$\sqrt{S}$\\\hline
		Type-I,X&$125.09$&$[126;\,1000]$&$[60;\,1000]$&$[80;\,1000]$&$[2;\,20]$&$[0.95;\,1]$&$ m_H^2 s_\beta c_\beta$&$14\ $TeV\\\hline
		Type-II,Y&$125.09$&$[500;\,1000]$&$[500;\,1000]$&$[580;\,1000]$&$[2;\,20]$&$[0.95;\,1]$&$ m_H^2 s_\beta c_\beta$&$14\ $TeV\\\hline
	\end{tabular}
		\caption{The input parameters for the $h-$ scenario, the masses are in GeV.}
		\label{scanh}
\end{table}
In connection with this issue, we present the loop-induced kappa trick in the $h-$scenario for the two final states $\gamma\gamma$ and $\gamma Z$, 
\begin{equation}
\kappa_{\gamma\gamma}^2 = \frac{\Gamma^{2HDM}(h\to \gamma\gamma)}{\Gamma^{SM}(h\to \gamma\gamma)}\label{kappa_hgaga_Zga},\qquad  
\kappa_{\gamma Z}^2= \frac{\Gamma^{2HDM}(h\to \gamma Z)}{\Gamma^{SM}(h\to \gamma Z)}
\end{equation}and the corresponding results are presented in figures \ref{fig:kappa-trick_h1} and \ref{fig:kappa-trick_h2}, for the four types of 2HDMs. Hence, in the Type-I, (figure \ref{fig:kappa-trick_h1}, upper panel), $\kappa_{\gamma\gamma}$ and $\kappa_{\gamma Z}$ can deviate substantially from unity for smaller charged Higgs boson masses due to the charged-Higgs loop contribution to the $h\gamma\gamma$ and $h\gamma Z$ coupling. This contribution can have a significant positive or negative sign in Type-I and Type-X (figure \ref{fig:kappa-trick_h2}, upper panel), respectively  , while in Type-II (figure \ref{fig:kappa-trick_h1}, lower panel) and Type-Y (figure \ref{fig:kappa-trick_h2}, lower panel), large contributions are always negative and suppress both $\kappa_{\gamma\gamma}$ and $\kappa_{\gamma Z}$. The large positive deviations in the Type-I are $12\%(5\%)$ for $\kappa_{\gamma\gamma}(\kappa_{\gamma Z})$ and $9\%(4\%)$ for the Type-X, while they are of the order of $1\%$ or less in Type-II and Y. Moreover, the negative deviations can reach $-12\%(-7\%)$ in Type-I, $-3\%(-1.02\%)$ in the Type-II, $-12\%(-7\%)$ in Type-X, and $-3\%(-1.\%)$ for Type-Y. From these results, it's evident that the deviations on the $\kappa_{\gamma\gamma}$ are all the time much larger than in $\kappa_{\gamma Z}$, and that can be explained by the fact that the sensitivity of charged loop is much larger in the $h\gamma\gamma$ coupling than in $h\gamma Z$ one. 

It is worth mentioning that the large deviations in the Type-I is set for charged Higgs masses $m_{H^\pm} \leq 690\,$ \gev\, irregularly with the $\tan\beta$ and $\cos(\beta-\alpha)$ parameters. Moreover, the $\kappa_{\gamma\gamma}(\kappa_{\gamma Z})$ enhancement becomes sizable $\geq 10\%(3\%)$ for $m_{H^\pm} \in [87.2,\ 100]$ \gev,\ $\cos(\beta-\alpha)\in [0.06,0.14]$ and $\tan\beta\geq 10$, while the suppression are important $\leq -10\%(-6\%)$ for $\cos(\beta-\alpha)\geq 0.27$, $m_{H^\pm}\in[182.3,\ 680]([128.4,\ 689.1])\ $\gev\ and  $\tan\beta\in [3,\ 8.6]([3.0,\ 19.2])$. 

\noindent
For Type-II, it's obvious that the parameter space is rigorously constrained : $\cos(\beta-\alpha)\leq 0.04$, $\tan\beta \leq 8$, leading to $\sin(\beta-\alpha)\sim 1$, and as the deviation is of kind's suppression, then it is coming from the bottom quark and the charged Higgs loop contributions which remained vulnerable here as $m_{H^\pm} \geq 580\, $\gev\ in such model's type.

\noindent
The Type-X (Y) are, theoretically, similar to the Type-I (II), but under the different constraints applied to the parameter space, the results can be much different for these types. For instance, in Type-X, deviations are possible for the full parameter space. The large enhancement is of $9\%(4\%)$ for $\kappa_{\gamma\gamma}(\kappa_{\gamma Z})$ is possible for $\cos(\beta-\alpha) \sim 0.1$, $m_{H^\pm} \sim 100\ $\gev\ and $\tan\beta \sim 17$, moreover, important $\kappa_{\gamma\gamma}(\kappa_{\gamma Z})$ enhancement  $ \geq 5\%(2\%)$ are possible for $\cos(\beta-\alpha)\leq 0.10$, $\tan\beta\geq 17$ and a charged Higgs masse $m_{H^\pm} \in [98.8,\ 155]([98.8,\ 126])\ $\gev,\ while the suppression are possible in general, irrespective the values of charged Higgs mass and $\cos(\beta-\alpha)$. Unlike, large $\kappa_{\gamma\gamma}(\kappa_{\gamma Z})$ suppression $ \leq -10\%(-5\%)$ favorites a  charged Higgs masses $m_{H^\pm} \in [170.8, 644]([165.6,\ 650])\,$ \gev, a large $\cos(\beta-\alpha)\in [0.27,0.31]([0.25,\ 0.31])$ and $\tan\beta\in [5.7,\ 7.1]([3.4,\ 8.1])$.

As far as the Type-Y, the enhancement of $0.67\%(0.25\%)$ for $\kappa_{\gamma\gamma}(\kappa_{\gamma Z})$ is possible for $\cos(\beta-\alpha)\leq 0.02$ and independently of $\tan\beta$ and $m_{H^\pm}$, which is owed to the bottom quark loop, while the
\begin{figure}[H]
	\begin{minipage}{0.5\textwidth}
		\centering
		\includegraphics[height=4.8cm,width=6.8cm]{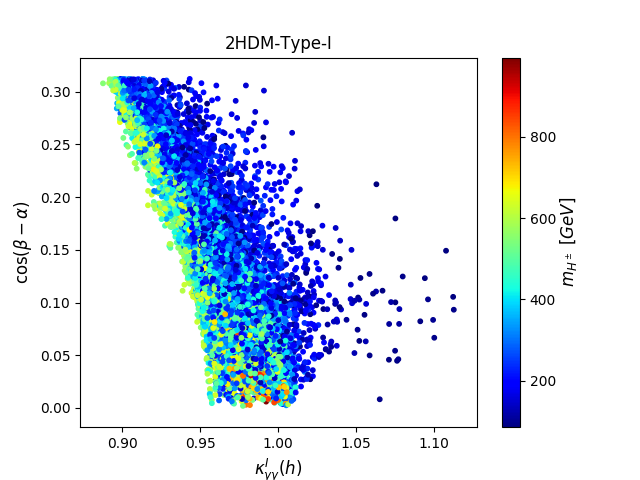}
	\end{minipage}
	\begin{minipage}{0.5\textwidth}
		\centering
		\includegraphics[height=4.8cm,width=6.8cm]{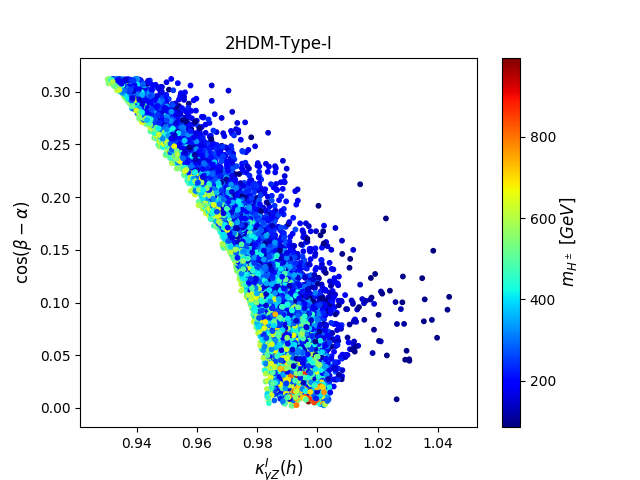}
	\end{minipage}
	\begin{minipage}{0.5\textwidth}
		\centering
		\includegraphics[height=4.8cm,width=6.8cm]{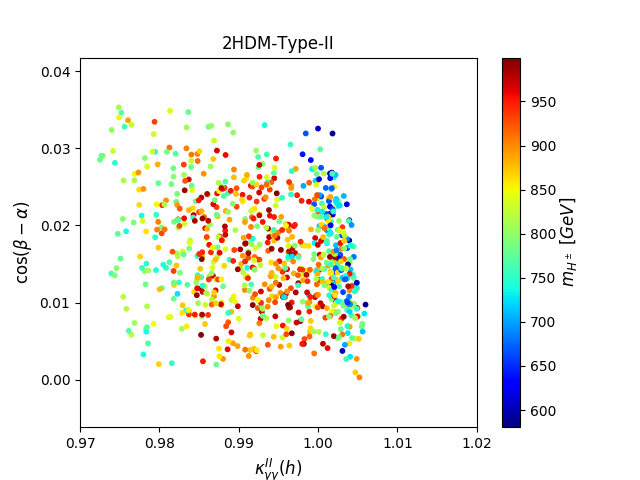}
	\end{minipage}
	\begin{minipage}{0.5\textwidth}
		\centering
		\includegraphics[height=4.8cm,width=6.8cm]{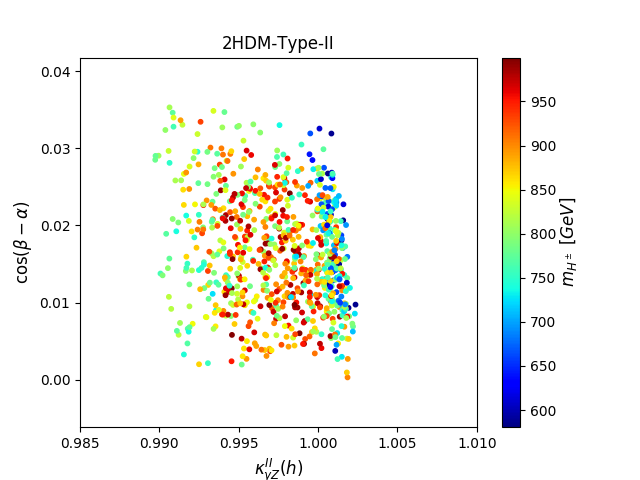}
	\end{minipage}
	\caption{The $h\gamma\gamma$(left panel) and $h\gamma Z$(right panel) $\kappa$-trick for $h-$scenario in the types I(upper panel) and II(lower panel) of 2HDMs as a function of $\cos(\beta-\alpha)$ and $m_{H^\pm}$ (\gev).}
	\label{fig:kappa-trick_h1}
\end{figure}
\begin{figure}[H]
	\begin{minipage}{0.5\textwidth}
		\centering
		\includegraphics[height=4.8cm,width=6.8cm]{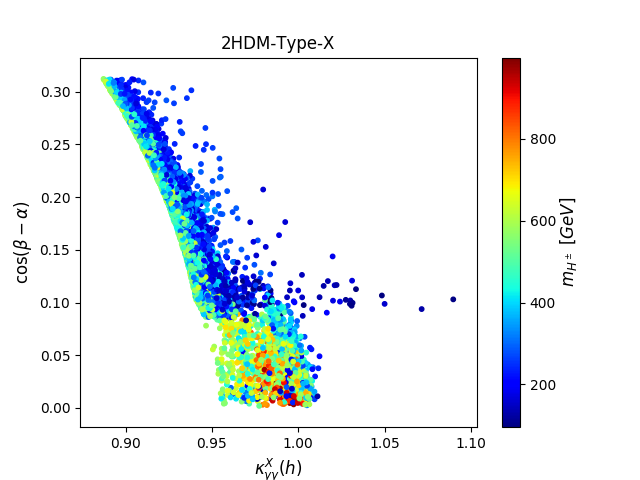}
	\end{minipage}
	\begin{minipage}{0.5\textwidth}
		\centering
		\includegraphics[height=4.8cm,width=6.8cm]{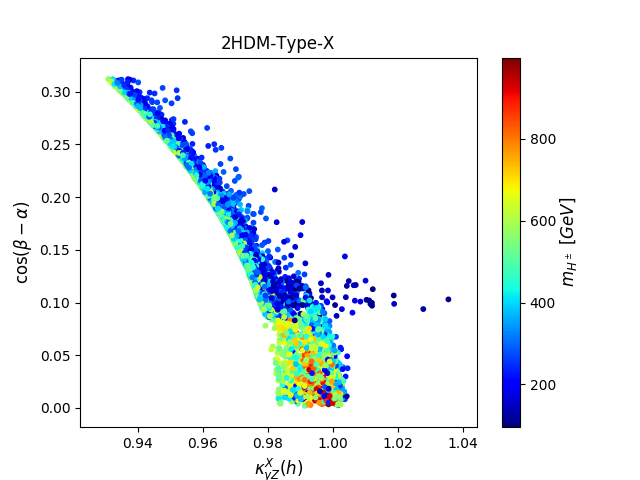}
	\end{minipage}
	\begin{minipage}{0.5\textwidth}
		\centering
		\includegraphics[height=4.8cm,width=6.8cm]{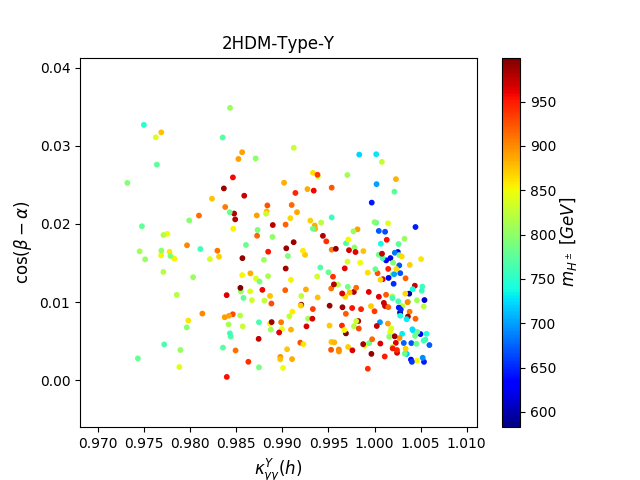}
	\end{minipage}
	\begin{minipage}{0.5\textwidth}
		\centering
		\includegraphics[height=4.8cm,width=6.8cm]{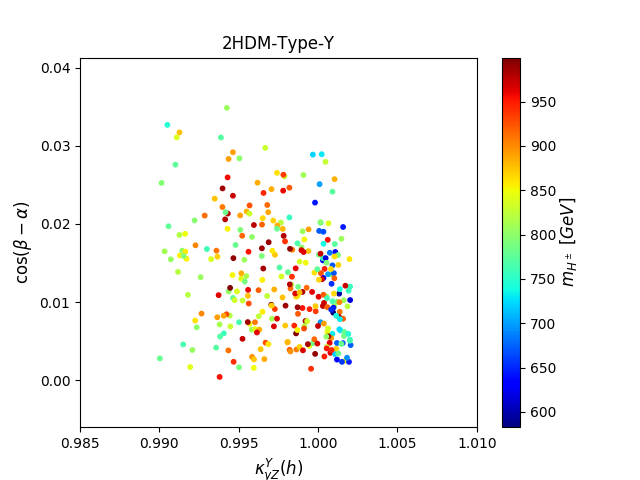}
	\end{minipage}
	\caption{The $h\gamma\gamma$ (left panel) and $h\gamma Z$ (right panel) $\kappa$-trick for $h-$scenario in the types X(upper panel) and Y(lower panel) of 2HDMs as a function of $\cos(\beta-\alpha)$ and $m_{H^\pm}$ (\gev).}
	\label{fig:kappa-trick_h2}
\end{figure}
\noindent
the suppression demands $m_{H^\pm} > 750\, $\gev\ and the $\tan\beta\in [2.5,12]$ regardless the allowed values of $\cos(\beta-\alpha)$.

The signal strengths in Eq.(\ref{eq:widthBSM}) can be rewritten as:
\begin{equation}
\mu_{\gamma\,V}(h)=\frac{[\sigma(pp \to h)\times BR(h \to \gamma V)]^{\rm 2HDM}}{[\sigma(pp \to h)\times BR(h \to \gamma V)]^{\rm SM}}
\label{eq:sigstrength-2hdm}
\end{equation}
The involved cross-section is computed at LO, NLO, and NNLO by SusHi\footnote{The public tool Sushi-1.7.0} \cite{Harlander:2012pb}, using the following LHAPDF set : MMHT2014lo68cl, MMHT2014nlo68cl, MMHT2014nnlo68cl, and MMHT2014nnlo68cl. The di-photon and $Z$-photon branching ratios, as well as the single Higgs production cross-section in the SM, are given in Table \ref{sm_xsbr} .
{\renewcommand{\arraystretch}{1.1} 
{\setlength{\tabcolsep}{0.2cm}
\begin{table}[H]
	\centering
	\begin{tabular}{|c|c|c|c|c|c|}
		\hline	
		observables&$BR(h\to \gamma Z$)&$BR(h\to\gamma\gamma)$&$\sigma_{pph}^{LO}/pb$&$\sigma_{pph}^{NLO}/ pb$&$\sigma_{pph}^{NNLO}/ pb$\\\hline
		SM values&$1.54\times10^{-3}$&$2.27\times 10^{-3}$&$16.22327$&$38.54630$&$51.4482460$\\\hline
	\end{tabular}
	\caption{The $h \to \gamma Z, \gamma\gamma $ branching ratios, and the production cross-sections in pb in the SM.}
	\label{sm_xsbr}
\end{table}
				
\noindent
For a deep understanding of the variations of the signal strengths $\mu_{\gamma \gamma}$ and  $\mu_{\gamma Z }$, we analyze the cross-section and decay branching ratio separately, for this reason, we have begun by studying the coupling or kappa trick as they are responsible for variations.

In the SM, the gluon fusion process with the internal top quark dominates the production cross-section. This holds in general for the four types of 2HDMs, but as $\tan\beta$ grows, the contribution from the gluon fusion process with internal bottom quarks may become more substantial. In order to understand qualitatively the impact of the different production rates in Fig \ref{fig:cross-section}. The gluon-gluon fusion cross-section could be expressed as follows:
\begin{align}
\sigma_{NNLO}^{2HDM}(g g\to h) = \kappa_u^2 C_{tt}+\kappa_u \kappa_d C_{tb}+\kappa_d^2 C_{bb}
\end{align}

\noindent
where $C_{tt}, C_{bb}$ and $C_{tb}$ are the contributions from top-quark, bottom-quark, and top-bottom interference. In the SM, $\kappa_u = \kappa_d = 1$, hence, $\sigma_{NNLO}^{SM} (gg \to h)$ $\approx C_{tt}$. The production signal strength could be explained as follows:
\begin{align}
\mu_{gg} &= \frac{\sigma_{NNLO}^{2HDM}(gg\to h)}{\sigma_{NNLO}^{SM}(gg\to h)}
=\kappa_u^2\left(1+\frac{\kappa_d\times C_{tb}}{\kappa_u\times C_{tt}}+\frac{\kappa_d^2\times C_{bb}}{\kappa_u^2\times C_{tt}}\right)
\end{align}

\noindent
In the 2HDMs Type-I and X, the $\kappa_u = \kappa_d $, thus the signal strengths of the production rate are as follows:
\begin{align}
\mu_{gg}^{\rm 2HDM-I,X} = \kappa_u^2\left(1+\frac{C_{tb}}{C_{tt}}+\frac{C_{bb}}{C_{tt}}\right),
\end{align}
while in the 2HDMs Type-II and Y, the signal strengths of the production rate takes the following form:
\begin{align}
&\mu_{gg}^{\rm 2HDM-II,Y}\approx \kappa_u^2 \left(1+\frac{\kappa_d\times C_{tb}}{\kappa_u\times C_{tt}}+\frac{\kappa_d^2\times C_{bb}}{\kappa_u^2\times C_{tt}}\right)\\
&= \kappa_u^2 \left(1-\tan\alpha\tan\beta\frac{C_{tb}}{C_{tt}}+\tan^2\alpha\tan^2\beta\frac{C_{tb}}{C_{tt}}\right)
\end{align}
If we take into account the fact that $C_{bb}<<|C_{tb}|<<C_{tt}$, thus $\frac{C_{bb}}{C_{tt}}<<\frac{|C_{tb}|}{C_{tt}}<< 1$. As a result, the production signal strengths are reduced to converge approximately to the top's reduced coupling to take the form below: 
\begin{align}
&\mu_{gg}^{\rm 2HDM-I,X} \approx \kappa_u^2\left(1+\frac{C_{tb}}{C_{tt}}\right)\approx \kappa_u^2\approxeq \kappa_t^2\label{mup_IX}\\
&\mu_{gg}^{\rm 2HDM-II,Y}\approx \kappa_u^2 \left(1-\tan\alpha\tan\beta\frac{C_{tb}}{C_{tt}}\right)\approx \kappa_u^2\approxeq \kappa_t^2\label{mup_IIY}
\end{align}
In Figure \ref{fig:cross-section}, we show the production signal strengths for the four types of 2HDMs; Type-I (upper left), Type-II (upper right), Type-X (lower left) and Type-Y (lower right). This figure, as well as Eqs. (\ref{coupling-h1}-\ref{coupling-h2}) and (\ref{mup_IX}-\ref{mup_IIY}), shows that these signal strengths of production can truly replicate the $\kappa_{t}^2$ in the four types of 2HDMs. The four types of the 2HDMs show the same behavior, and the types I (II) are closely similar to the types X (Y) with a slight difference coming from the applied constraints. \textcolor{black}{
In both types, the predicted signal strengths of production at $95\%$ CL are consistent with its experimental measurement given in Table \ref{experimental_signal_strenghts}.}
\begin{figure}[H]
	\begin{minipage}{0.5\textwidth}
		\centering
		\includegraphics[height=4.8cm,width=6.8cm]{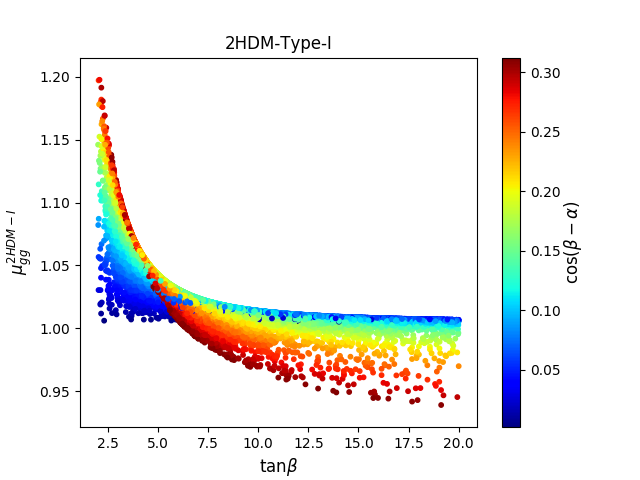}
	\end{minipage}
	\begin{minipage}{0.5\textwidth}
		\centering
		\includegraphics[height=4.8cm,width=6.8cm]{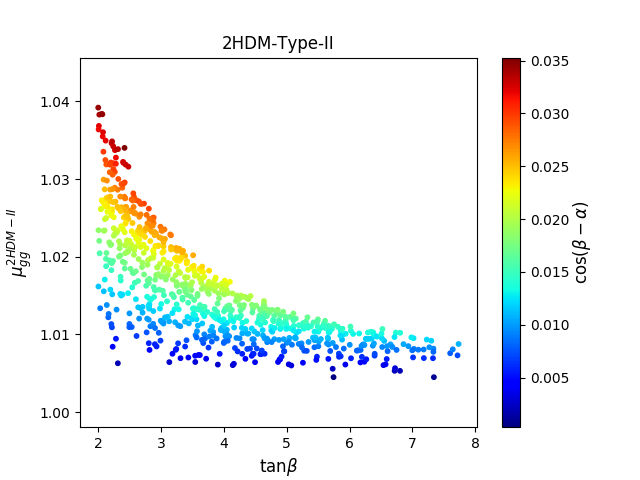}
	\end{minipage}
	\begin{minipage}{0.5\textwidth}
		\centering
		\includegraphics[height=4.8cm,width=6.8cm]{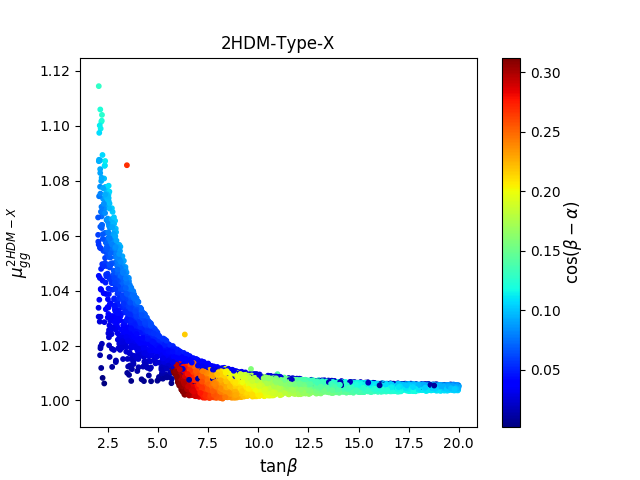}
	\end{minipage}
	\begin{minipage}{0.5\textwidth}
		\centering
		\includegraphics[height=4.8cm,width=6.8cm]{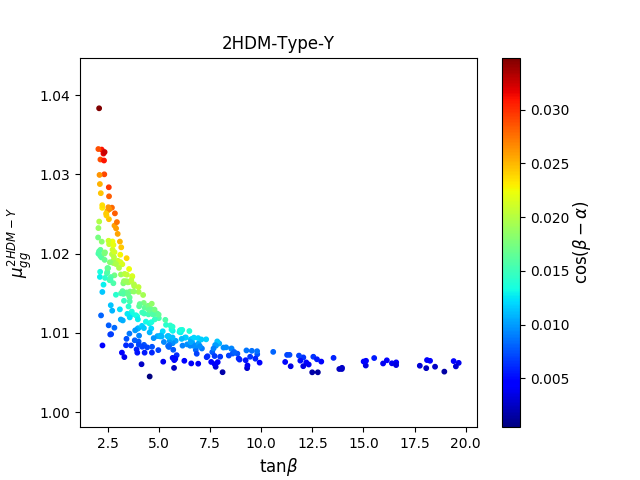}
	\end{minipage}
	\caption{$\mu_{gg}^{2HDM}$ at $95\%$ CL as a function of $\tan\beta$ vs. $\cos(\beta-\alpha)$, in Type-I(upper left), Type-II(upper right), Type-X(lower left) and Type-Y(lower right) of 2HDMs.}
	\label{fig:cross-section}
\end{figure}	
\noindent
Regarding the corresponding partial widths $\Gamma(h \to \gamma\,V)$, they are given by:
\begin{align}
&\Gamma(h \to \gamma\gamma)=\frac{G_F \alpha^2 m_h^3}{128\sqrt{2}\pi^3}\bigg|\sum_{f}\kappa_{f}Q_f^2 N_c A_{1/2}(\tau_f)+ \kappa_W A_{1}(\tau_W)+\kappa_{hH^{\pm}H^{\mp}}A_{0}(\tau_{H^\pm})\bigg|^2\\
&\Gamma(h \to \gamma Z)=\frac{G_F \alpha^2 m_h^3}{64\sqrt{2}\pi^3}\bigg(1-\frac{m_Z^2}{m_h^2}\bigg)^{3}\bigg|\sum_{f}\kappa_f Y_F^f + \kappa_W Y_W + \kappa_{hH^\pm H^\mp}Y_{H^\pm}\bigg|^2
\end{align}
\noindent
Figures \ref{fig:br-correlation-thdm1} and \ref{fig:br-correlation-thdm2} depicts the branching ratios of $h \to \gamma \gamma$ and $h \to \gamma Z$ normalized to their SM values in four different types of 2HDMs. As it can be seen, it's clear that the enhancement and suppression may occur in the four types of model, and as we know that the charged loop and the gouge boson contribution are type-independent, as a results, the difference between the four types comes from the fermionic contribution and how the constraints applied affect the parameter space.

In Type-I (figure \ref{fig:br-correlation-thdm1}, upper panel), the deviations on the normalized branching ratio are of, $\gamma\gamma\in [-30\%,16\%]$, $\gamma Z\in [-25\%,3\%]$. Were the $\gamma\gamma(\gamma Z)$ enhancement are only possible for the smaller charged Higgs masses $m_{H^\pm}\leq 184(100)$ \gev\ and $cos(\beta-\alpha)\in [0,\ 0.21]([0,\ 0.14])$, $\tan\beta > 7.6(10.6)$, while the suppression are possible for the full allowed parameter space, moreover, the large suppression $\leq -20\%$ take place for $m_{H^\pm} \in [140,\ 721]([160,\ 680])$  \gev, $\tan\beta\in [2,\ 9.2]([2,\ 3.8])$ and far from alignment limit $\cos(\beta-\alpha) > 0.12(0.19)$. \textcolor{black}{In the region closest to the exact alignment limit i.e.,} $\cos(\beta-\alpha)\sim 0$ and $m_{H^\pm} >200$ \gev\ the branching ratio get suppressed by a $-13\sim -4.3\%$.

In the Type-II (figure \ref{fig:br-correlation-thdm1}, lower panel), the deviations are relatively smaller compared to the Type-I; $\gamma\gamma\in [-8\%,7\%], \gamma Z\in [-7\%,7\%]$, which can be explained by the fact that the charged and top contributions interfere destructively with the contributions of $W$-Boson, also as $\cos(\beta-\alpha)\sim 0$, then we approx the exact alignment limit, as a result, the deviations of branching ratio relative to the SM values would be minimal. However, the $\gamma\gamma(\gamma Z)$ enhancement are permitted in the full allowed parameter space, while the suppression requires $m_{H^\pm}\geq 600$ \gev.

\textcolor{black}{Types X (Y) in Figure \ref{fig:br-correlation-thdm2} support the same explanations given for Types I (II) with slight differences generated by the different constraints applied in these types.}

\begin{figure}[H]
	\begin{minipage}{0.5\textwidth}
		\centering
		\includegraphics[height=4.8cm,width=6.8cm]{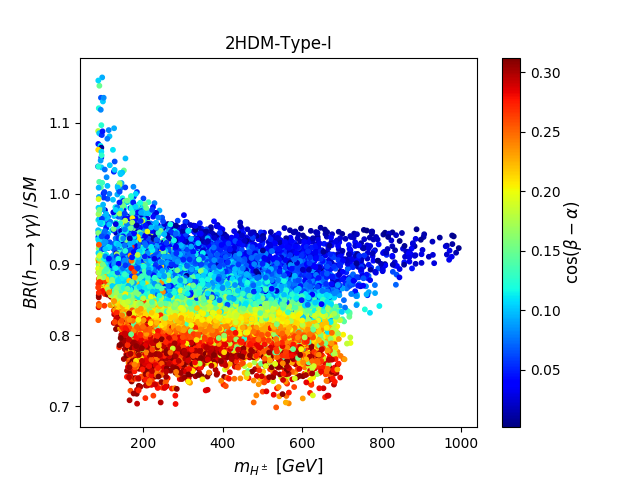}
	\end{minipage}
	\begin{minipage}{0.5\textwidth}
		\centering
		\includegraphics[height=4.8cm,width=6.8cm]{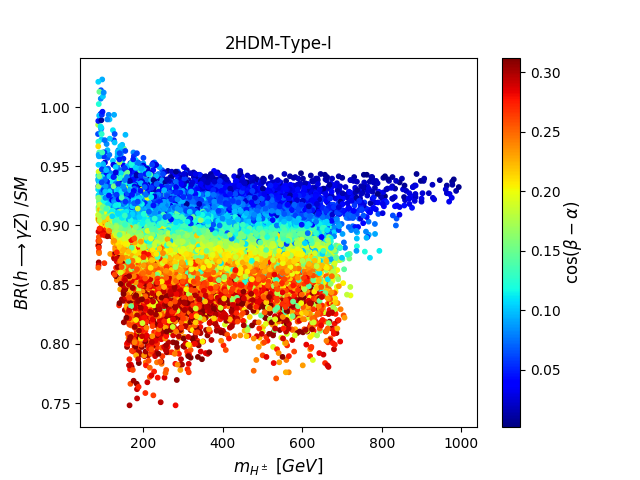}
	\end{minipage}
	\begin{minipage}{0.5\textwidth}
		\centering
		\includegraphics[height=4.8cm,width=6.8cm]{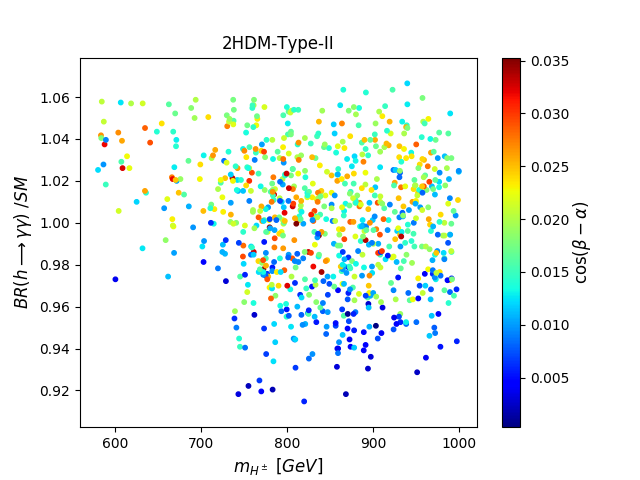}
	\end{minipage}
	\begin{minipage}{0.5\textwidth}
		\centering
		\includegraphics[height=4.8cm,width=6.8cm]{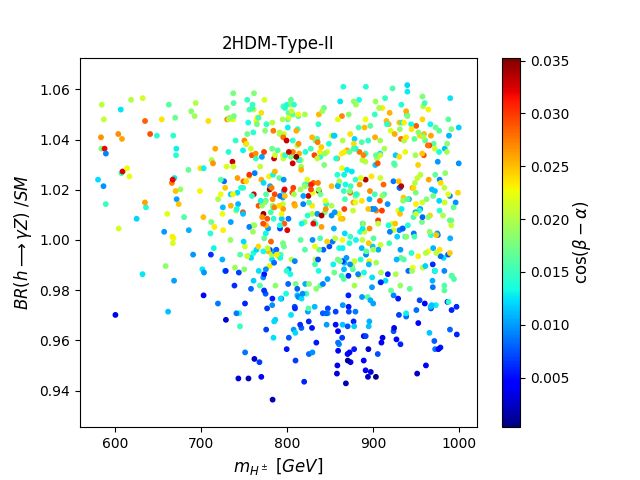}
	\end{minipage}
	\caption{Normalize $BR_{\gamma\gamma}(h)$(left panel) and $BR_{\gamma Z}(h)$(right panel) as a function of $m_{H^\pm}$ (\gev) and $\cos(\beta-\alpha)$ in the 2HDMs type-I(upper panel) and II(lower panel) : $m_h = 125.09\ $GeV.}
	\label{fig:br-correlation-thdm1}
\end{figure}

\begin{figure}[H]
	\begin{minipage}{0.5\textwidth}
		\centering
		\includegraphics[height=4.8cm,width=6.8cm]{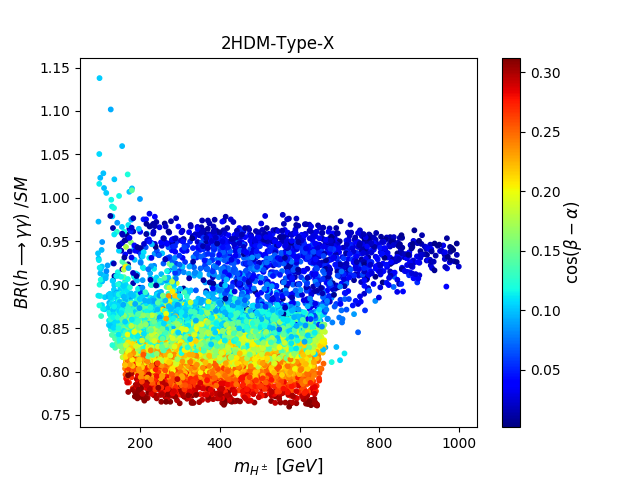}
	\end{minipage}
	\begin{minipage}{0.5\textwidth}
		\centering
		\includegraphics[height=4.8cm,width=6.8cm]{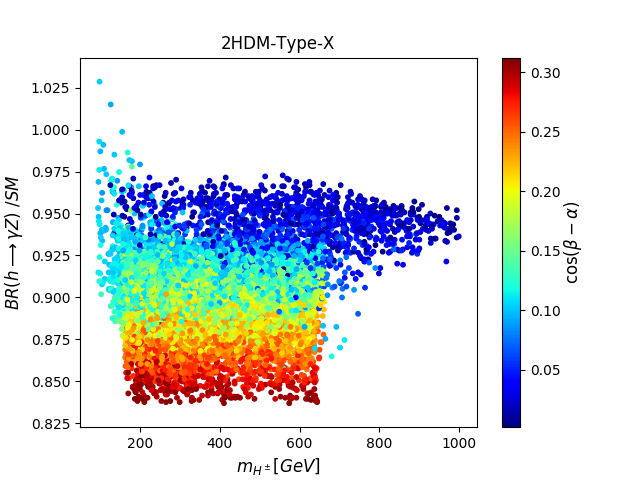}
	\end{minipage}
	\begin{minipage}{0.5\textwidth}
		\centering
		\includegraphics[height=4.8cm,width=6.8cm]{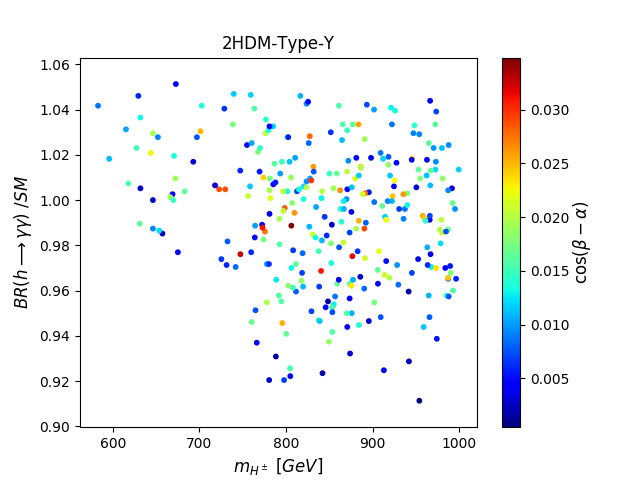}
	\end{minipage}
	\begin{minipage}{0.5\textwidth}
		\centering
		\includegraphics[height=4.8cm,width=6.8cm]{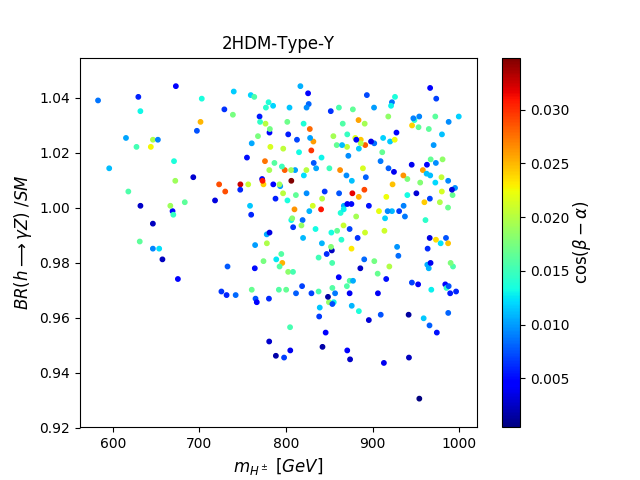}
	\end{minipage}
	\caption{Normalize $BR_{\gamma\gamma}(h)$(left panel) and $BR_{\gamma Z}(h)$(right panel) as a function of $m_{H^\pm}$ (\gev) and $\cos(\beta-\alpha)$ in the 2HDMs Type-X(upper panel) and Y(lower panel): $m_h = 125.09\ $GeV.}
	\label{fig:br-correlation-thdm2}
\end{figure}

We note that the decay channels $h\to \gamma\gamma$ and $h\to \gamma Z$ have identical loop diagrams either in the SM or beyond. Therefore, one could expect the correlation between their decays or branching ratio to behave like a linear relation in the four types of 2HDMs.
The Figure \ref{fig:br-correlation-thdm}, show the correlations of the branching ratios present in the figure \ref{fig:br-correlation-thdm2} as a function of $\cos(\beta-\alpha)$ for the four types of 2HDMs; Type-I (upper left), Type-II (upper right), Type-X (lower left) and Type-Y (lower right). The normalized $h\to\gamma\gamma$ and $h\to \gamma Z$ branching ratios to the SM ones are presented on the x-axis and y-axis, respectively. From these figures, it's clear that these normalized branching ratios are effectively correlated positively, and the Type-I(and II) are very similar to the Type-X(and Y). In most regions of 2HDMs Type-I and X parameters space, the $BR(h\to\gamma \gamma)$ and $BR(h\to\gamma Z)$ are in deficit with respect to SM prediction, while in the types II and Y, the two regions where the 2HDMs predictions are in excess or in deficit are moderated.

In the Type-I, the overlap region where the normalized $BR(h\to\gamma \gamma)> 1$ and $BR(h\to\gamma Z) > 1$ is very narrowed and require $\cos(\beta-\alpha)\leq 0.14$, $\tan\beta > 10$ and a light-charged Higgs boson, i.e., $m_{H^\pm}\in [87.2,\ 100.8]$ \gev,\ while the overlap suppressed region; i.e, $BR(h\to \gamma\gamma) < 1$ and $BR(h\to \gamma Z) < 1$ is not constrained and dominate the allowed parameter space. In 2HDMs Type-II, the region with  $BR(h\to\gamma \gamma)> 1$ and $BR(h\to\gamma Z) > 1$ are allowed by the whole allowed parameter space, while the opposite case $BR(h\to \gamma\gamma) < 1$ and $BR(h\to \gamma Z) < 1$ requires $m_{H^\pm} > 600\ $\gev. 

In Type X, we fall in region were $BR(h\to\gamma \gamma)> 1$ and $BR(h\to\gamma Z) > 1$ for $\cos(\beta-\alpha) < 0.14$, $\tan\beta\in [11.8,\ 19.2]$ and a light-charged Higgs boson; i.e $m_{H^\pm}\in [98.18,\ 180.2]\ $\gev,\  while the region were $BR(h\to\gamma \gamma) < 1$ and $BR(h\to\gamma Z) < 1$ is the dominant and requires $m_{H^\pm} > 95\ $\gev. For the Type-Y, the region in excess $BR(h\to\gamma \gamma)> 1$ and $BR(h\to\gamma Z) > 1$ is possible in the whole allowed parameter space, while the deficit region $BR(h\to\gamma \gamma) < 1$ and $BR(h\to\gamma Z) < 1$ require $m_{H^\pm} > 630$ \gev. 				
\begin{figure}[H] 
	\begin{minipage}{0.5\textwidth}
		\centering
		\includegraphics[height=4.8cm,width=6.8cm]{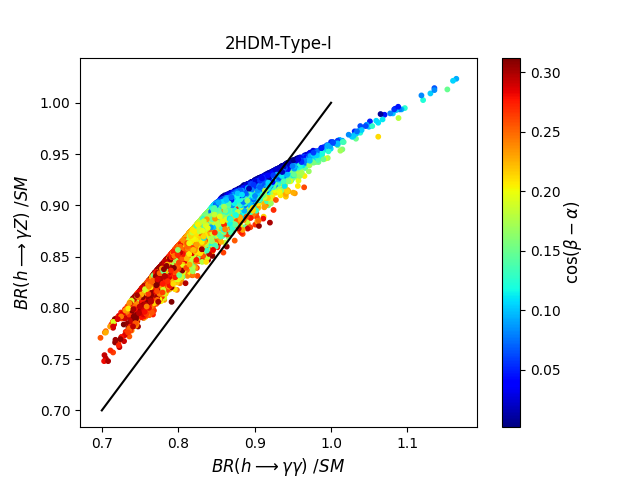}
	\end{minipage}
	\begin{minipage}{0.5\textwidth}
		\centering
		\includegraphics[height=4.8cm,width=6.8cm]{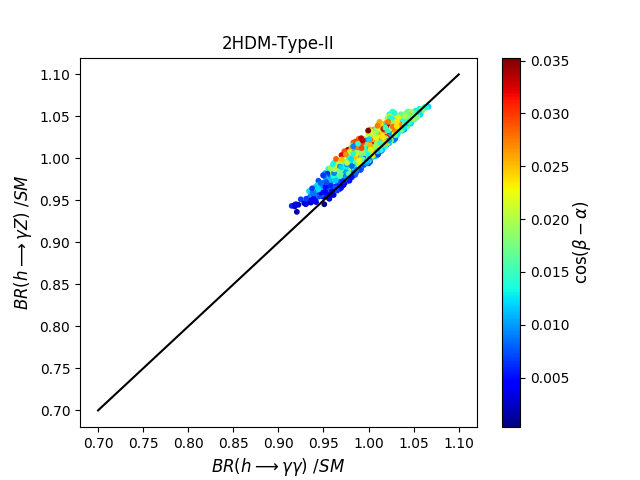}
	\end{minipage}
	\begin{minipage}{0.5\textwidth}
		\centering
		\includegraphics[height=4.8cm,width=6.8cm]{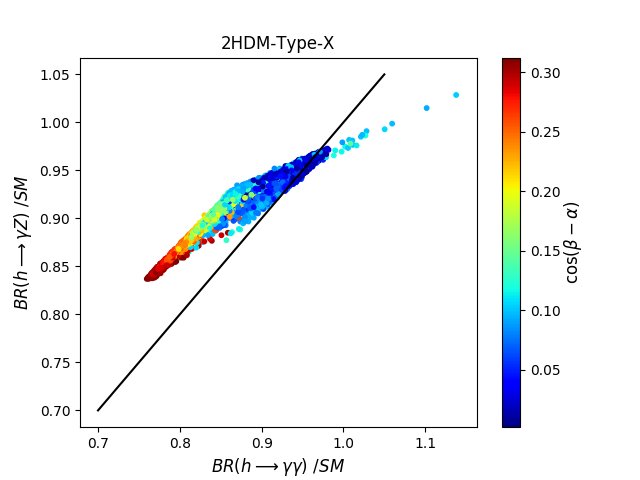}
	\end{minipage}
	\begin{minipage}{0.5\textwidth}
		\centering
		\includegraphics[height=4.8cm,width=6.8cm]{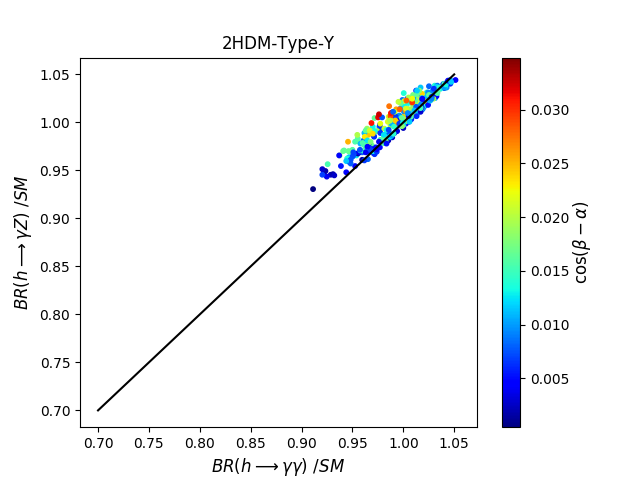}
	\end{minipage}
	\caption{Correlation between normalized $BR_{\gamma\gamma}(h)$ and $BR_{\gamma Z}(h)$ in types I(upper left), II(upper right), X(lower left) and Y(lower right) of 2HDMs : $m_h = 125.09\ $GeV. The coding shows $\cos(\beta-\alpha)$. The diagonal black line shows the $BR(h\to \gamma Z)/SM = BR(h\to \gamma\gamma)/SM$. }
	\label{fig:br-correlation-thdm}
\end{figure}
\noindent
In view of the foregoing, the signal strengths in Eq.(\ref{eq:sigstrength-2hdm}) shall consist as
\begin{align}
\mu_{\gamma\,V}&
= \mu_{gg}^{2HDM}\times\frac{BR^{2HDM}(h \to \gamma\,V)}{BR^{SM}(h \to \gamma\,V)}
\approxeq \kappa_t^2\times \frac{BR^{2HDM}(h \to \gamma\,V)}{BR^{SM}(h \to \gamma\,V)}
\end{align}	

and the variations in the kappa-trick analyzed above have an immediate impact on the cross-section and branching ratio, thus on the signal strengths of the SM-like Higgs boson. We look at the four types of CP-conserving 2HDM separately since the results are heavily influenced by the fermion coupling structure. Figure \ref{fig:signal-strength}, shows the correlation of signal strengths $\mu_{\gamma\gamma}$ and $\mu_{\gamma Z}$ in the four types of \thdm s; Type-I (upper left), Type-II (upper right), Type-X (lower left) and Type-Y (lower right).
In the Type-I, we found that $\mu_{\gamma\gamma}\in [0.76,1.16]$ and $\mu_{\gamma Z}\in [0.82,1.02]$, thus the $\mu_{\gamma\gamma}$ are more affected by the correction than $\mu_{\gamma Z}$. That's explained by the fact that in the four different types of 2HDMs, the $h\to \gamma Z$ decay is much less sensitive to the charged scalar loops than the $h\to \gamma\gamma$ decay.

It is obvious that the charged Higgs loop contribution to the  $h \to \gamma Z$ and $h\to \gamma\gamma$ accounts for a considerable part of the deviations in $\mu_{\gamma\gamma}$ and $\mu_{\gamma Z}$. A significant charged Higgs contribution or the presence of a light pseudo-scalar boson, below the $h\to AA$ threshold, which raises the SM-like Higgs total width, might cause substantial deviations from unity even for $\cos(\beta-\alpha)\approx 0$. In contrast, the charged Higgs loop is small in the limit where $m_{H^\pm}^2 >> v^2$, as a result the $h\gamma\gamma$ and $hZ\gamma$ couplings are mostly dictated by the top and bottom loops relative sizes to the $W$-loop, which enters with the opposite sign. 
In the Type-II, $\mu_{\gamma\gamma}\in [0.92,1.07]$ and $\mu_{\gamma Z}\in [0.94,1.07]$, the signal strengths can be driven by both the top and bottom quark couplings as it's clear from the previous figure, also the charged loop contribution is small as $m_{H^\pm} > 580$ \gev \  in our analysis, moreover $\mu_{\gamma\gamma}$ and $\mu_{\gamma Z}$ can be enhanced and suppressed in the decoupling regime as a result of the suppression of the total $h$ width. The amount of possible suppression on the $\mu_{\gamma\gamma}$ and $\mu_{\gamma Z}$ decreases systematically with decreasing of $\cos(\beta-\alpha)$. The charged Higgs loop's contribution to the $h\to \gamma Z$ and $h\to \gamma\gamma$ amplitude causes another effect to manifest in signal strengths $\mu_{\gamma Z}$ and, $\mu_{\gamma \gamma}$. In particular, the $\kappa_{hH^\pm H^\mp}\sim \frac{-2m_{H^\pm}}{v}$ produces a steady non-decoupling contribution that suppresses the such amplitude for a range of intermediate charged Higgs masses.
\begin{figure}[H]
	\begin{minipage}{0.5\textwidth}
		\centering
		\includegraphics[height=4.8cm,width=6.8cm]{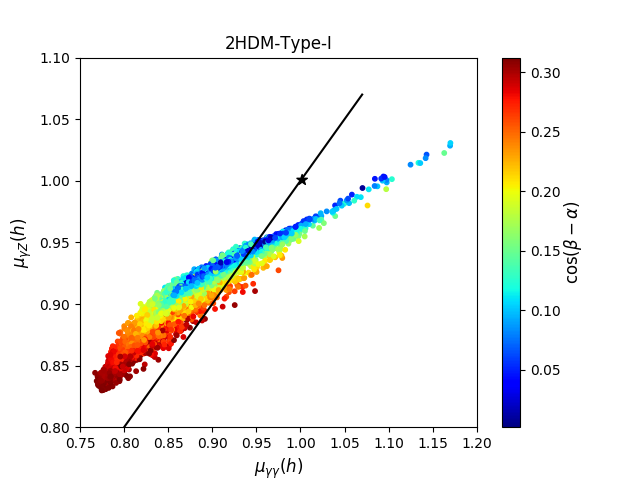}
	\end{minipage}
	\begin{minipage}{0.5\textwidth}
		\centering
		\includegraphics[height=4.8cm,width=6.8cm]{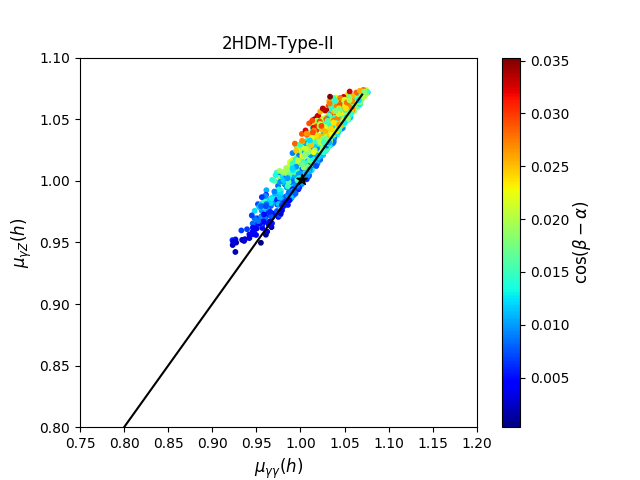}
	\end{minipage}
	\begin{minipage}{0.5\textwidth}
		\centering
		\includegraphics[height=4.8cm,width=6.8cm]{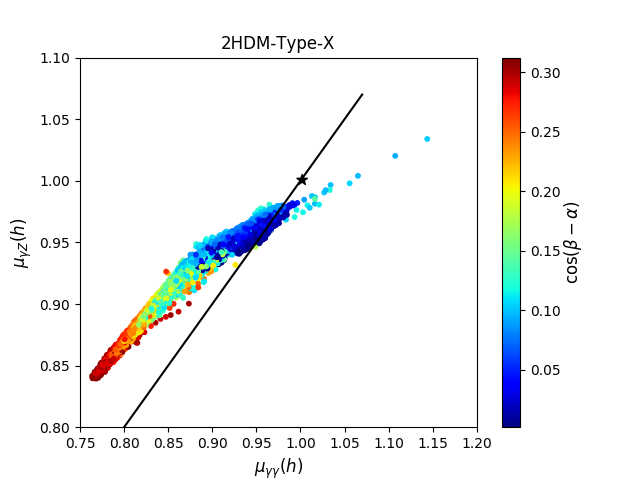}
	\end{minipage}
	\begin{minipage}{0.5\textwidth}
		\centering
		\includegraphics[height=4.8cm,width=6.8cm]{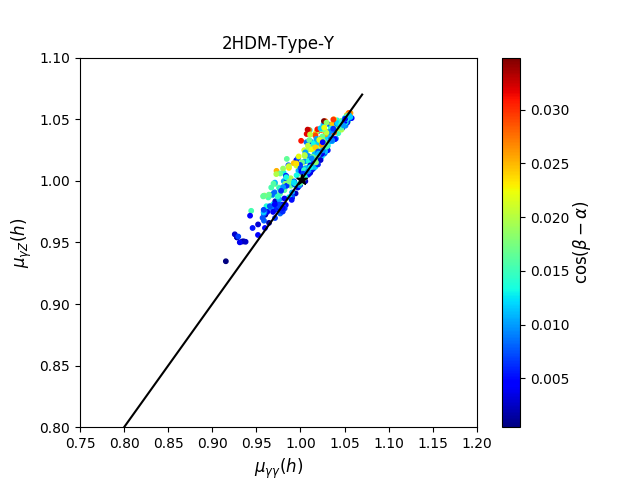}
	\end{minipage}
	\caption{Correlation between $\mu_{\gamma\gamma}(h)$ and $\mu_{\gamma Z}(h)$ vs. $cos(\beta-\alpha)$ at NNLO in types I(upper left), II(upper right), X(lower left) and Y(lower right) of 2HDMs. The diagonal black line shows $\mu_{\gamma Z }=\mu_{\gamma \gamma}$, while the asterisk is for SM values.}
	\label{fig:signal-strength}
\end{figure}
The types X and Y are similar to types I and II, respectively.  In order to understand what’s implied by the deviations in the signal strengths on the 2HDMs model structure, we analyzed the enhancement and suppression in the four versions of 2HDMs separately. In type I, the enhancement on the $\mu_{\gamma\gamma}$ take place for: $m_{H^\pm}\in [86.5,\ 200.7]$ \gev, $\cos(\beta-\alpha)\in [0,\ 0.21]$ and $\tan\beta \in [7.6,\ 19.8]$, while for $\mu_{\gamma Z}$, it's taking place for $m_{H^\pm}\in [87.2,\ 113.3]$ \gev, $\cos(\beta-\alpha)\in [0,\ 0.14]$ and $\tan\beta \in [8.1,\ 19.8]$. The suppression are possible in the whole parameter space either for $\mu_{\gamma\gamma}$ and $\mu_{\gamma Z}$. In Type II, enhancement is possible across the entire parameter space as long as the suppression demand is $m_{H^\pm} > 600$ \gev\ and $\cos(\beta-\alpha) < 0.02(0.016)$ for $\mu_{\gamma\gamma}(\mu_{\gamma Z})$ regardless of $\tan\beta$. For Type X, the $\mu_{\gamma \gamma}$ enhancement demand, $m_{H^\pm}\in [98.1,\ 200.1]$ \gev, $\cos(\beta-\alpha)\in [0, 0.14]$ and $\tan\beta \in [11.8, 19.2]$, whereas the $\mu_{\gamma Z}$ enhancement has a strong influence on the $m_ {H^\pm} \in [98.8, 155.5]$ \gev, $cos(\beta-\alpha)\in [0.9, 0.1]$ and $\tan\beta\in [17.6, 19.2]$. For both $\mu_{\gamma \gamma}$ and $\mu_ {\gamma Z}$, the suppression demand, $m_{H^\pm}$ to be greater than $95$ \gev. For the Type Y, the enhancement in both $\mu_{\gamma \gamma}$ and $\mu_{\gamma Z}$ remains unrestricted by the parameter space, while the suppression demand $m_{H^\pm} \geq 646$ \gev, and $\cos(\beta-\alpha) < 0.025 (0.018)$ for $\mu_{\gamma\gamma} (\mu_{\gamma Z})$ regardless of $\tan\beta$

We note that the observations of the $h \to \gamma Z$ decay are significantly more difficult than observing the $h \to \gamma\gamma $ decay from an experimental standpoint. The highest limit on the production cross-section times the branching ratio for the signal $pp\to h\to \gamma Z$ is only roughly seven times the SM forecast according to the most recent LHC data. 
\subsection{In the \idm}
\label{sec:idm-res}  
\textcolor{black}{For the IDM, we focused on scenarios in which $H$ is only a subdominant DM candidate and does not account for $100\%$ of dark matter in the universe. The signal strengths is reduced to,}
\begin{align}
\mu_{\gamma\,V} = \frac{BR^{\rm IDM}(h \to \gamma\,V)}{BR^{\rm SM}(h \to \gamma\,V)}
\end{align}
where the narrow width approximation was employed in this case. The widths of decays are as follows:
\begin{align}
&\Gamma^{\rm IDM}(h \to \gamma\gamma)=\frac{G_F\alpha^2 m_h^3}{128\sqrt{2}\pi^2}\bigg|\frac{4}{3}A_{1/2}(\tau_t)+A_1(\tau_W)+ \kappa_{hH^{\pm}H^{\mp}} A_0(\tau_H^\pm)\bigg|^2\\
&\Gamma^{\rm IDM}(h\longrightarrow \gamma Z)=\frac{G_F^2\alpha}{64\pi^2}m_W^2 m_h^2\left(1-\frac{m_Z^2}{m_h^2}\right)^3\bigg|Y_t + Y_W + \kappa_{hH^{\pm}H^{\mp}} Y_H^\pm\bigg|^2
\end{align}
We note that the partial widths of the tree-level $h$ decay into SM particles and the loop-mediated decay into gluon-gluon in the \idm are identical to the corresponding ones in the SM. The only decay rates of $h$-SM that are modified in the \idm compared to the SM are those for the $h \to \gamma\gamma$ and $h\to \gamma Z$ processes. Consequently, the branching ratios and the signal strengths of these decays channels can deviate with respect to the SM ones by two additional contributions: the first comes from the charged scalar exchanged in loops gives an extra contribution to the $h\gamma\gamma$ and $h\gamma Z$ amplitude, while the last coming
from the invisible decay $h\to HH, AA$ to the total decay width $\Gamma(h)$ when $m_H<m_h/2$ and/or $m_A<m_h/2$. Therefore, the $h\to \gamma\gamma$ and $h \to \gamma Z$ channels are perfect ones to reveal some information about the extra scalars present in this model.

To start with, we distinguish between two cases: whether the $m_{H}<m_{A},m_{H^{\pm}}$ or the inert states are degenerate $m_{H^\pm} = m_A = m_H+1\,$\gev. If the former, we set the following inputs and we perform a random scan over the IDM parameter space; see table ~\ref{idm_gen}, taking into account the theoretical and experimental constraints discussed above.
\begin{table}[H]
	\centering
	\begin{tabular}{|c|c|c|c|c|c|c|}
		\hline
		Parameters&$m_h$&$m_H$&$m_A$&$m_{H^\pm}$&$\lambda_2$&$\mu_2^2$(GeV$^2$)\\\hline
		range&$125.09$&$[10;\, 1000]$&$[10;\, 1000]$&$[10;\, 1000]$&$[0;\, 4\pi/3]$&$[-10^{6};\, 10^{-6}]$\\\hline	
	\end{tabular}
	\caption{IDM inputs parameter for the H-scenario, ~$m_{H}<m_{A},m_{H^{\pm}}$, the masse are in GeV.}\label{idm_gen}
\end{table}

\noindent
In Fig.\ref{fig:idm-1}, we show the prediction of the signal strengths $\mu_{\gamma \gamma} $ and $\mu_{\gamma Z}$ in the IDM framework. In the upper panel, we depict both $\mu_{\gamma\gamma}$(left) and $\mu_{\gamma Z}$(right) as functions of dark matter candidate's mass $m_H$ and $\lambda_L\sim \kappa_{hHH}$(this is taken into account while calculating the WMIP relic density).  The lower panel depict the same quantity as a function of the charged Higgs boson mass $m_{H^\pm}$ and $\lambda_3\sim \kappa_{hH^{\pm}H^{\mp}}$.

By analyzing the data of these figures, we found that the enhancement on the $\mu_{\gamma\gamma} > 1$ is possible in general for,
\begin{align*}
m_H\in [62.94,164.93]\ \textrm{GeV},\ m_{H^\pm}\in [79.46, 186.2]\ \textrm{GeV},\  \lambda_3\in [-0.77,-0.1]\ \rm{and} \ \lambda_L\in [-0.48,-0.01]
\end{align*}

\noindent
while for the $\mu_{\gamma Z}$ the enhancement is not possible due the experimental  constraints from Xenon1T.  The enhancement on the $\mu_{\gamma \gamma}$ is only possible in the region where the invisible decays are closed $m_H > 62.5$ \gev, and that's because when the invisible channels are open, both $\mu_{\gamma Z}$ and $\mu_{\gamma \gamma}$ are suppressed, as they are damped by the invisible decays widths. Furthermore, we found that the suppression are possible for both $\mu_{\gamma \gamma}$ and $\mu_{\gamma Z}$ in the whole region of the parameter space. The difference between $\mu_{\gamma \gamma}$ and $\mu_{\gamma Z}$ can be explained by the fact that, the decay $h\to \gamma\gamma$  is much more sensible  to the charged-loop contributions than in $h\to \gamma Z$ decay. The $\mu_{\gamma \gamma}$  reaches its best enhancement, which is $8.52\%$ for $m_H = 63.06$ \gev, $m_{H^\pm} = 117.06$ \gev, $\lambda_L=-0.42$ and $\lambda_3 = -0.42$, while the $\mu_{\gamma Z}$
is limited in $-0.45\%$ as a maximum for, $m_H\in [62.06,\ 63.15]$ \gev, $m_{H^\pm}\in [92.69, 117.49]$ \gev, $\lambda_3\in [-0.53,-0.28]$ and  $\lambda_L\in [-0.42,-0.21]$.

\begin{figure}[H] 
	\begin{minipage}{0.5\textwidth}
		\centering
		\includegraphics[height=4.8cm,width=6.8cm]{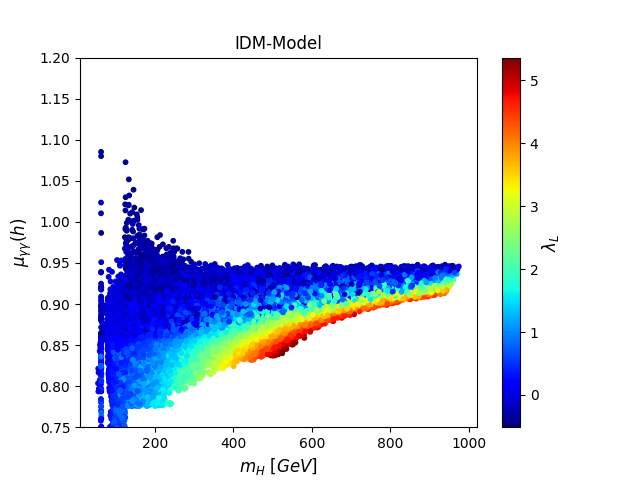}
	\end{minipage}
	\begin{minipage}{0.5\textwidth}
		\centering
		\includegraphics[height=4.8cm,width=6.8cm]{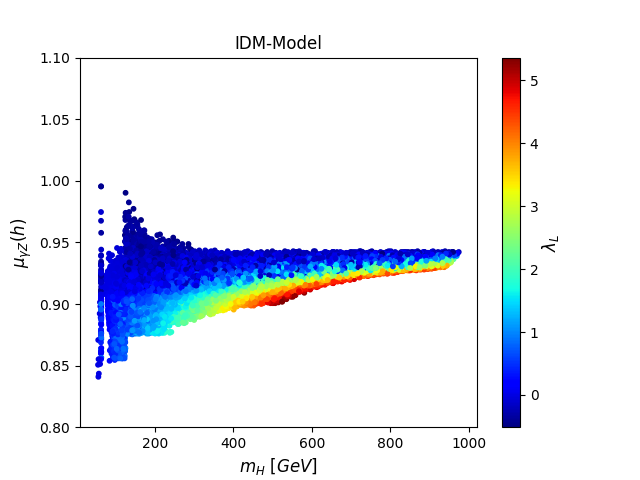}
	\end{minipage}
	\begin{minipage}{0.5\textwidth}
		\centering
		\includegraphics[height=4.8cm,width=6.8cm]{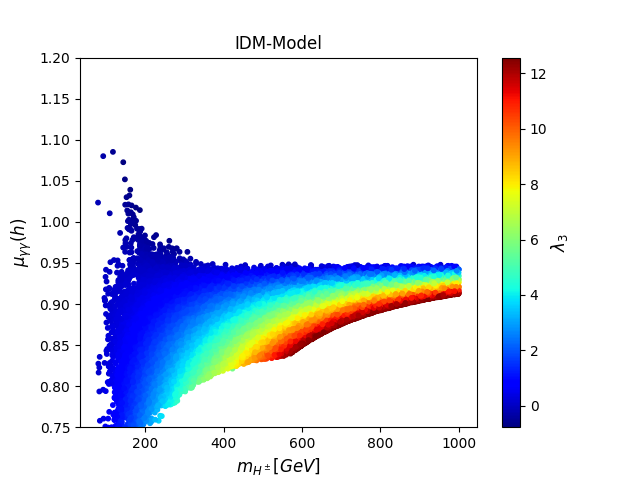}
	\end{minipage}
	\begin{minipage}{0.5\textwidth}
		\centering
		\includegraphics[height=4.8cm,width=6.8cm]{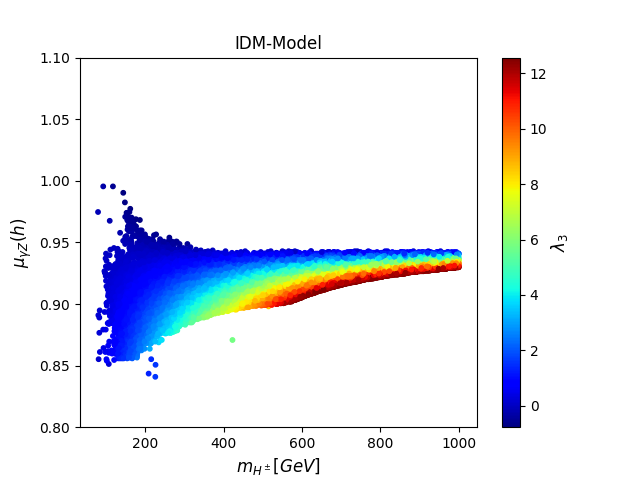}
	\end{minipage}
	\caption{signal strengths $\mu_{\gamma\gamma}(h)$(left panel) and $\mu_{\gamma Z}(h)$(right panel) in IDM(General case) as a function of $m_{H}$ overlaid on  $\lambda_L $ (upper panel), and $m_{H^\pm}$ overlaid on  $\lambda_3$ (lower panel).}
	\label{fig:idm-1}
\end{figure}

For the case of an approximately degenerate analysis of parameter space, we vary the inputs as set in Table \ref{idm_degen}.
\begin{table}[H]
	\centering
	\begin{tabular}{|c|c|c|c|c|c|c|}
		\hline
		Parameters&$m_h$&$m_H$&$m_A$&$m_{H^\pm}$&$\lambda_2$&$\mu_2^2$(GeV$^2$)\\\hline
		range&$125.09$&$[10;\, 1000]$&$m_H+1$&$m_H+1$&$[0;\, 4\pi/3]$&$[-10^{6};\, 10^{-6}]$\\\hline	
	\end{tabular}
	\caption{IDM inputs for the H-scenario, $m_{H^\pm} = m_A = m_H+1$. The masses are in GeV}\label{idm_degen}
\end{table}

\noindent
It is worth mentioning in this case that the results are slightly different compared with the general case, as the invisible decay is constrained by the LEP-II; i.e $m_{H^\pm} > 70$ \gev\ and also due to the Xenon1T experiment, which hardly constrained the parameter space in this case. The deviations in the signal strength are due to the contributions of the charged Higgs loop.

In Fig.\ref{fig:idm-2}, we show the results of signal strengths $\mu_{\gamma \gamma}$ and $\mu_{\gamma Z}$  in the same plan as in Fig.\ref{fig:idm-1} but this time with the approximately degenerate case as set in table \ref{idm_degen}. From these figures, it's obvious that the results show the same behavior as in Fig.\ref{fig:idm-1} except for the allowed parameter space, which is very small in this case. Moreover, we found with this setting that the $\mu_{\gamma \gamma}$ enhancement is possible for, $m_H\in [69.1,171.1]$ \gev, $\lambda_3 \in [-0.99,-0.04]$ and $\lambda_L \in [-0.5, -0.025]$.  For $\mu_{\gamma Z}$, the enhancement is possible for, $m_H\in [126.59,141.85]$ \gev, $\lambda_3 \in [-0.99,-0.71]$ and $\lambda_L \in [-0.5, -0.36]$. While the suppression are possible for both $\mu_{\gamma \gamma}$ and $\mu_{\gamma Z}$ in the whole region of allowed parameter space. The enhancements are bigger in the approximately degenerate case with respect to the general case and can reach $14.7\%$ for $\mu_{\gamma \gamma}$ and $1.62\%$ for $\mu_{\gamma Z}$ as maximum, while the suppression are small with respect to the former case  and, reach $-15.8(-10.19)\%$ for $\mu_{\gamma \gamma}(\mu_{\gamma Z})$ instead of $-27.08(-15.90)\%$ in the general case.
\begin{figure}[H] 
	\begin{minipage}{0.5\textwidth}
		\centering
		\includegraphics[height=4.8cm,width=6.8cm]{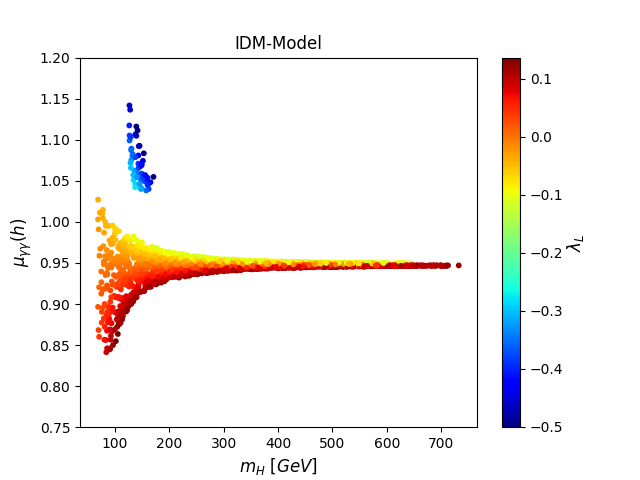}
	\end{minipage}
	\begin{minipage}{0.5\textwidth}
		\centering
		\includegraphics[height=4.8cm,width=6.8cm]{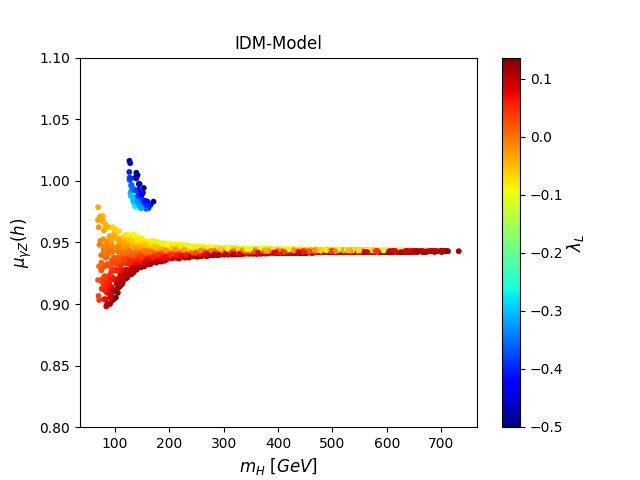}
	\end{minipage}
	\begin{minipage}{0.5\textwidth}
		\centering
		\includegraphics[height=4.8cm,width=6.8cm]{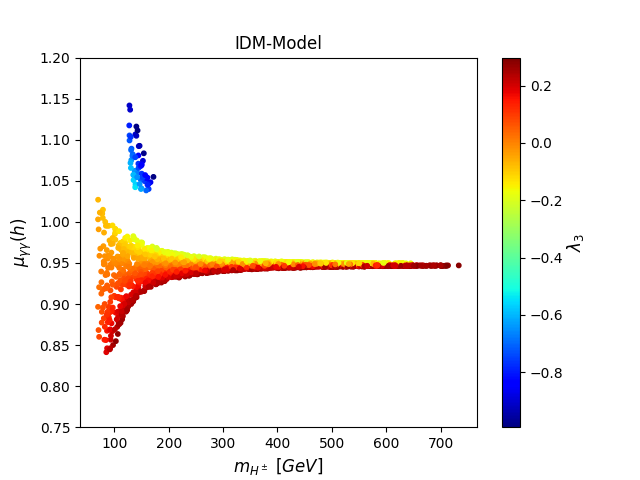}
	\end{minipage}
	\begin{minipage}{0.52\textwidth}
		\centering
		\includegraphics[height=4.8cm,width=6.8cm]{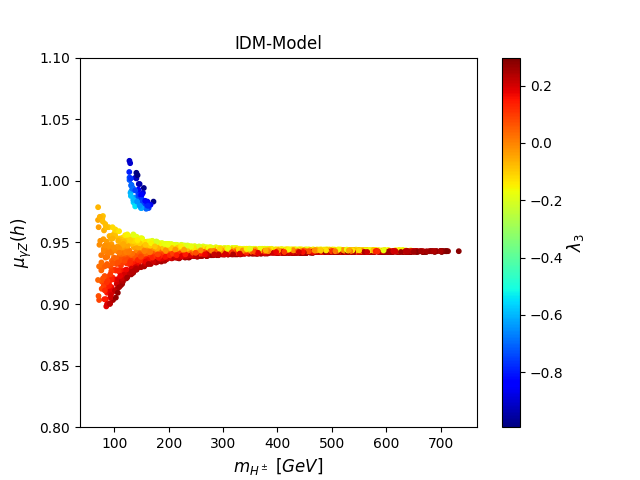}
	\end{minipage}
	\caption{signal strengths $\mu_{\gamma\gamma}(h)$(left) and $\mu_{\gamma Z}(h)$(right) in IDM (approximately degenerate case) as a function of $m_{H}$ overlaid on  $\lambda_L $ (upper panel), and $m_{H^\pm}$ overlaid on  $\lambda_3$ (lower panel).}
	\label{fig:idm-2}
\end{figure}
\noindent
Anyway, the \idm provides deviations on the $\mu_{\gamma Z}$ relatively small compared to $\mu_{\gamma\gamma}$. Additionally, we see that with additional data at the LHC, it would be feasible to determine whether the signal strengths is in excess, and with $\mu_{\gamma Z} > 1$ and $\mu_{\gamma \gamma} > 1$, the following scenario would be required: 
$62.9 < m_{H}(\rm{\gev}) < 171.1$ and, $\lambda_3 < 0\ \rm{and}\ \lambda_L < 0$,
on the other hand, if with this data, they found suppression  $\mu_{\gamma\gamma} <1 $ and $\mu_{\gamma Z} < 1$, then the following scenario will be favored : $\lambda_3 > 0\ \rm{and}\ \lambda_L >0$.

With the \idm\ in mind, we have examined the signal strengths correlation for both channels $\gamma\gamma$ and $\gamma Z$. The results are present in Figure \ref{fig:idm-3} for the two cases described above, namely : the general case (left panel) and the approximately degenerate case (right panel). From these figures, it's clear that the correlations of signal strengths are positive. Furthermore, in the general case (left), we found that $\mu_{\gamma Z} > \mu_{\gamma \gamma}$ in a large portion of the parameter space and that $\mu_{\gamma Z} < \mu_{\gamma \gamma}$ is only possible in the region where the invisible decays are closed $m_H> 62.9$ GeV and for $\lambda_3 \in [-0.77,\ 2.72]$. The region where $\mu_{\gamma Z} > \mu_{\gamma \gamma}$ requires $\lambda_3 > 0.026$ no matter the value of other parameters. For the approximately degenerate case, the situation is quite different as the constraints favorite the set of parameters where $\mu_{\gamma Z} < \mu_{\gamma \gamma}$. The situation were $\mu_{\gamma Z} > \mu_{\gamma \gamma}$ is only possible for $\lambda_3 \in [0.01,\ 0.27],\ \lambda_L \in [0.0,\ 0.13]$ and $m_H \leq 302$ GeV.  In general, for $\mu_{\gamma \gamma} > 1 $, where $W^\pm$ and $H^\pm$ loops contributions are constructive, we could have $\mu_{\gamma \gamma}\geq \mu_{\gamma Z}$. Instead, for  $\mu_{\gamma \gamma}\leq 1$, where $W^\pm$ and $H^\pm$ loops contributions are destructive, the situation were $\mu_{\gamma \gamma} \leq \mu_{\gamma Z}$ is the dominate and can be explained by the fact that $\mu_{\gamma \gamma}$ is more affected by the destructive interference between $W^\pm$ and $H^\pm$ than $\mu_{\gamma Z}$, which has a significantly larger contribution from $W^\pm$.

\noindent
In the general case, the overlap region where $\mu_{\gamma Z} > 1$ and where $\mu_{\gamma \gamma} > 1$  are not still survival as $\mu_{\gamma Z}$ is always suppressed. While for the approximately degenerate case, this overlap region of enhancement is much more important and is for $m_H \in [126.59,141.85]$ \gev, $\lambda_3 \in [-0.99, -0.71]$ and $\lambda_L \in [-0.50, -0.36]$. 

\begin{figure}[H] 
	\begin{minipage}{0.5\textwidth}
		\centering
		\includegraphics[height=4.8cm,width=6.8cm]{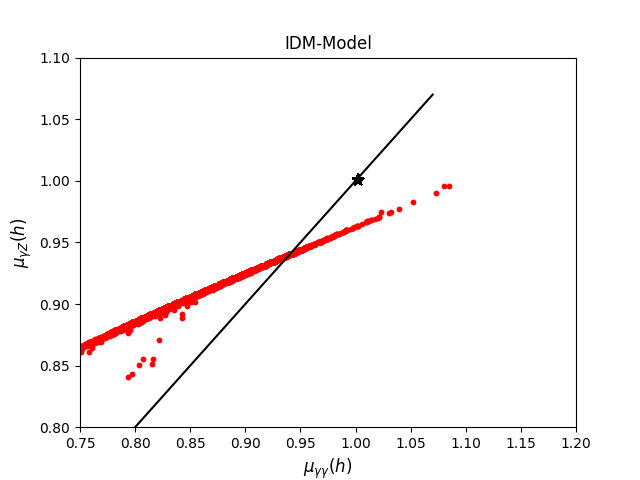}
	\end{minipage}
	\begin{minipage}{0.5\textwidth}
		\centering
		\includegraphics[height=4.8cm,width=6.8cm]{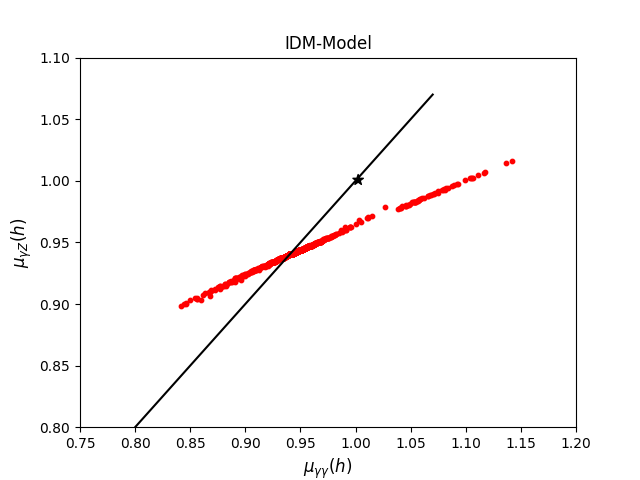}
	\end{minipage}
	\caption{Correlation between $\mu_{\gamma \gamma}(h)$ and $\mu_{\gamma Z}(h)$ in the IDM : general case (left), approximately degenerate case (right).}
	\label{fig:idm-3}
\end{figure}
By comparing the overlap region of enhancement to the single region where $\mu_{\gamma \gamma}$ and  $\mu_{\gamma Z}$ are enhanced, one can see that $\mu_{\gamma Z}$ can be enhanced if and only if $\mu_{\gamma \gamma}$ is enhanced, consequently, there is a positive correlation between $\mu_{\gamma Z}$ and $\mu_{\gamma \gamma}$, and a measurement yielding a different result would rule out this model.
\subsection{In the \htm}
\label{sec:htm-res}  
What further distinguishes the \htm is the existence of a newly doubly charged Higgs boson, $H^{\pm\pm}$, in addition to the simple one $H^{\pm}$. The decay width, and instead of using the expression given by Eq (\ref{eq:part-width-hgaZ-sm}), we will now have:
\begin{align}
\Gamma(h \to \gamma Z) &= {{\alpha\,G_F^2 m_{W}^2 m_{h}^3} \over {64\pi^2}} \bigg(1- {{m_Z^2} \over {m_h^2}} \bigg)^3 \bigg| \sum_{f} \kappa_{f} N_f {{Q_f \hat{v}_f}\over{c_W}} A_{1/2}^h(\tau_f,\lambda_f)+\kappa_{W} A_{1}^h(\tau_W,\lambda_W) + \nonumber\\
& \hspace{4.65cm} \sum_{S=H^\pm,H^{\pm\pm}} Q_S^2 \frac{m_W^2}{m_{S}^2} \kappa_{ZSS} \kappa_{hSS} A_{0}^h(\tau_S,\lambda_S) \bigg|^2.\nonumber\\
&= {{\alpha\,G_F^2 m_{W}^2 m_{h}^3} \over {64\pi^2}} \bigg(1- {{m_Z^2} \over {m_h^2}} \bigg)^3\bigg| Y_F + Y_W + Y_{H^\pm} + Y_{H^{\pm\pm}}\bigg|^2
\label{eq:part-width-hgaZ-htm}
\end{align}

\noindent
For better understanding the new physics contribution, we compute the squared moduli of each part between the absolute value brackets in Eq (\ref{eq:part-width-hgaZ-htm}), and for ease of notation, we refer by $X_W$, $X_t$, $X_{H^{\pm}}$ and $X_{H^{\pm\pm}}$ to the entire expression of the gauge boson $W$, top quark, single-charged Higgs boson and doubly charged Higgs boson, including all corresponding couplings. It appears, at first sight, that the terms may fairly differ by orders of magnitude as the quartic coupling $\lambda_1$ may be varied. Indeed, for $\lambda_1$ close to 0, like in the SM, we have that $|X_W|^2 \sim 3.19 \times 10^{2} |X_t|^2$ but it's more a $10^4$ times larger than $X_{H^{\pm}}$ and $X_{H^{\pm\pm}}$, as illustrated in the left panel of Figure \ref{fig:contri-inter-htm}. We also notice that the interference term contributions remain small (see right panel of Figure \ref{fig:contri-inter-htm}) so that the amplitude is arguably similar to the SM one, where the $W$ and top contributions are the most expressive.
On the other side, for non-vanishing $\lambda_1$, the situation has nearly changed. At the outset, the single-charged Higgs contribution is almost the same and is far below the top quark by two orders of magnitude, and whatever $\lambda_1>0\, {\rm or}\,<0$. Moreover, the corresponding interference term remains destructive.
\begin{figure}[H]
	\begin{minipage}{0.5\textwidth}
		\centering
		\includegraphics[height=4.8cm,width=6.8cm]{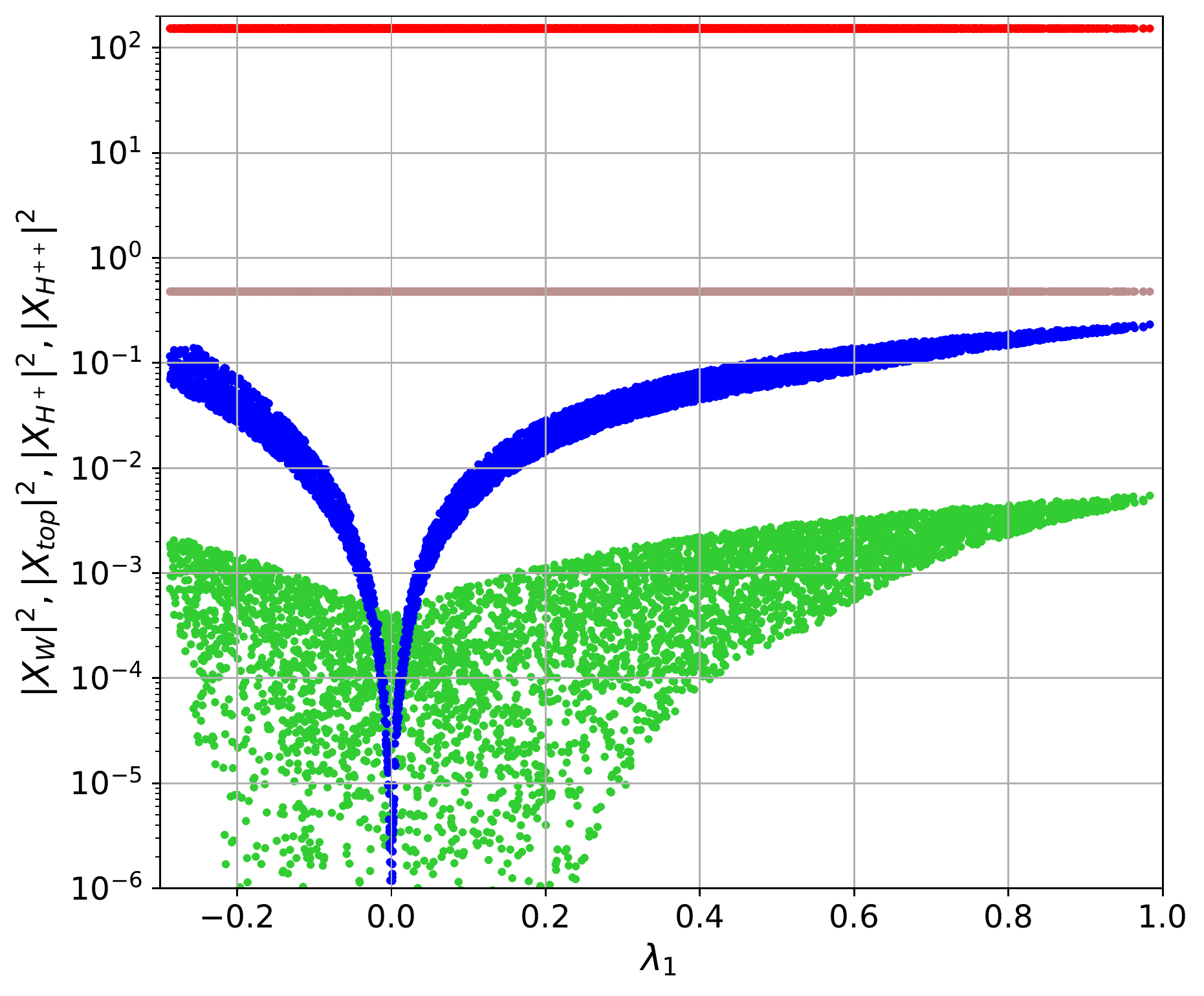}
	\end{minipage}
	\begin{minipage}{0.5\textwidth}
		\centering
		\includegraphics[height=4.8cm,width=6.8cm]{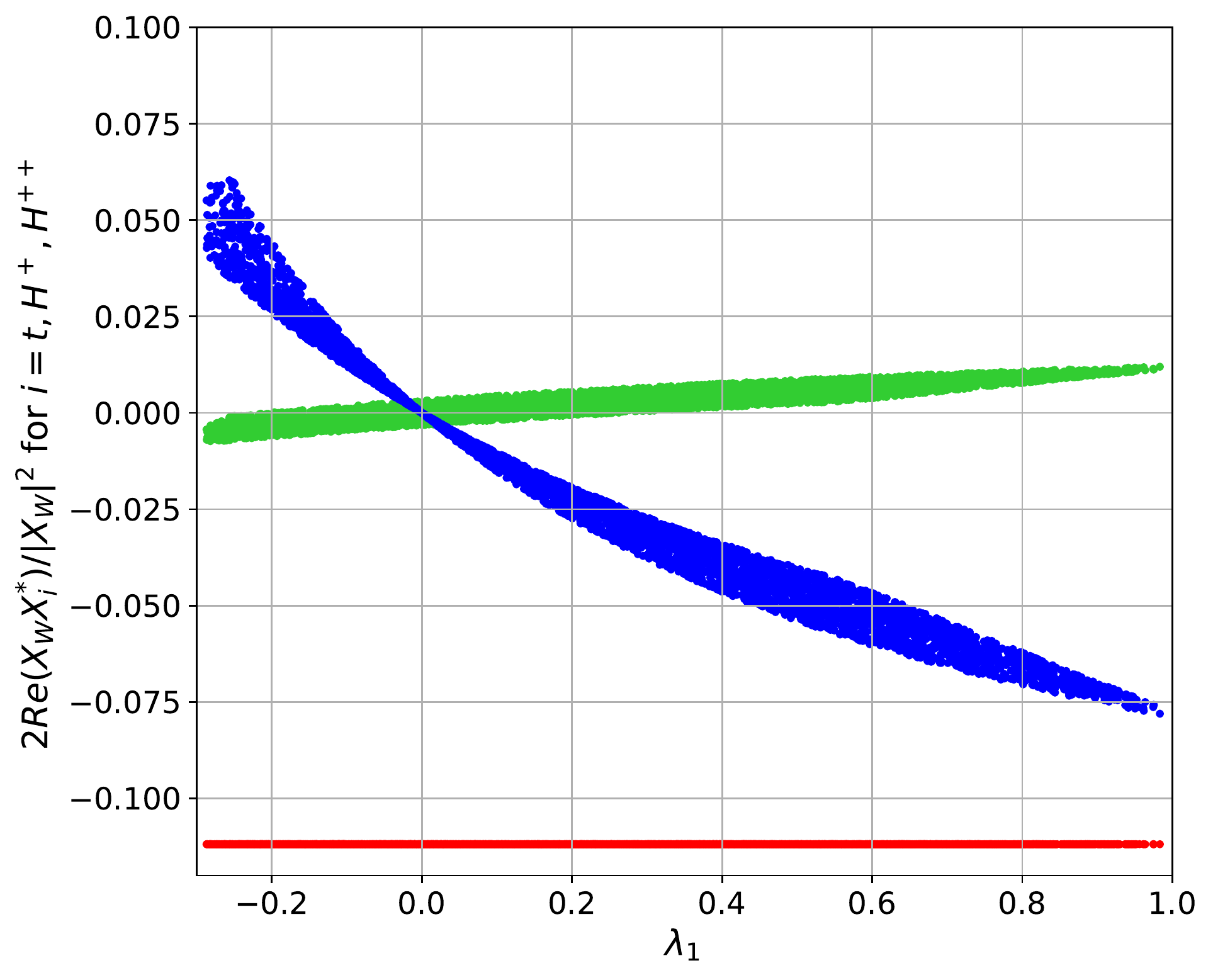}
	\end{minipage}
	\caption{(Left): the squared moduli of $W^\pm$ (red), top (brown), $H^{\pm}$ (green) and $H^{\pm\pm}$ (blue) contributions as a function of the coupling $\lambda_1$ in the framework of HTM. (Right): the interaction terms between $X_W$ and $X_t$ (red), $X_{H^{\pm}}$ (green) and $X_{H^{\pm\pm}}$ (blue) reduced to the squared modulus of $X_W$, as a function of the $\lambda_1$ coupling.}
	\label{fig:contri-inter-htm}
\end{figure}
\noindent
Quite the opposite in fact - the doubly charged Higgs contribution is non-negligible, being comparable to the top quark one either for $\lambda_1>0$ or $<0$. Not only that, the interference term $2\Re(X_W\,X_{H^{\pm\pm}}^{\ast})/|X_W|^2$ could be both constructive (for $\lambda_1<0$) and destructive (for $\lambda_1>0$). This latter underpins the importance of $\lambda_1$ sign, and thus those of the couplings $\kappa_{hH^{\pm}H^{\mp}}$ and $\kappa_{hH^{\pm\pm}H^{\mp\mp}}$, to probe the $\mu_{\gamma Z}$, for which we exhibit in Figure \ref{fig:correlation-htm} its correlation with the di-photon signal strength, $\mu_{\gamma\gamma}$. At this point, it is worth to mention that only when the $H^{\pm\pm}$ interference term is destructive that $\mu_{\gamma Z}(h)>1$ and so is $\mu_{\gamma\gamma}(h)>1$, which becomes more and more pronounced for light charged Higgs masses, i.e. $m_{H^{\pm}}\sim175$ (GeV) and $m_{H^{\pm\pm}}=150\sim160$ (GeV).
\begin{figure}[H]
	\begin{minipage}{0.5\textwidth}
		\centering
		\includegraphics[height=4.8cm,width=6.8cm]{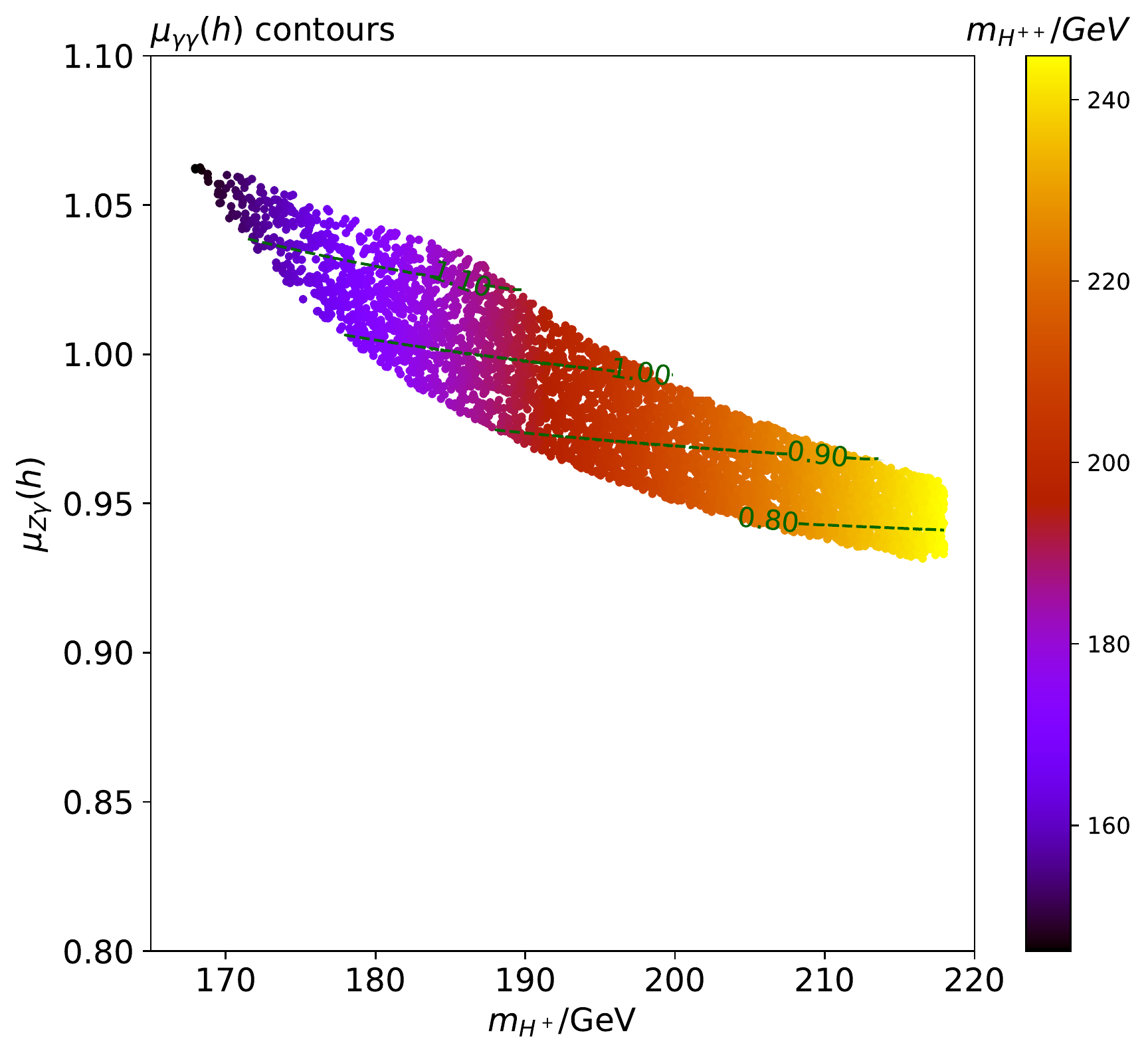}
	\end{minipage}
	\begin{minipage}{0.5\textwidth}
		\centering
		\includegraphics[height=4.8cm,width=6.8cm]{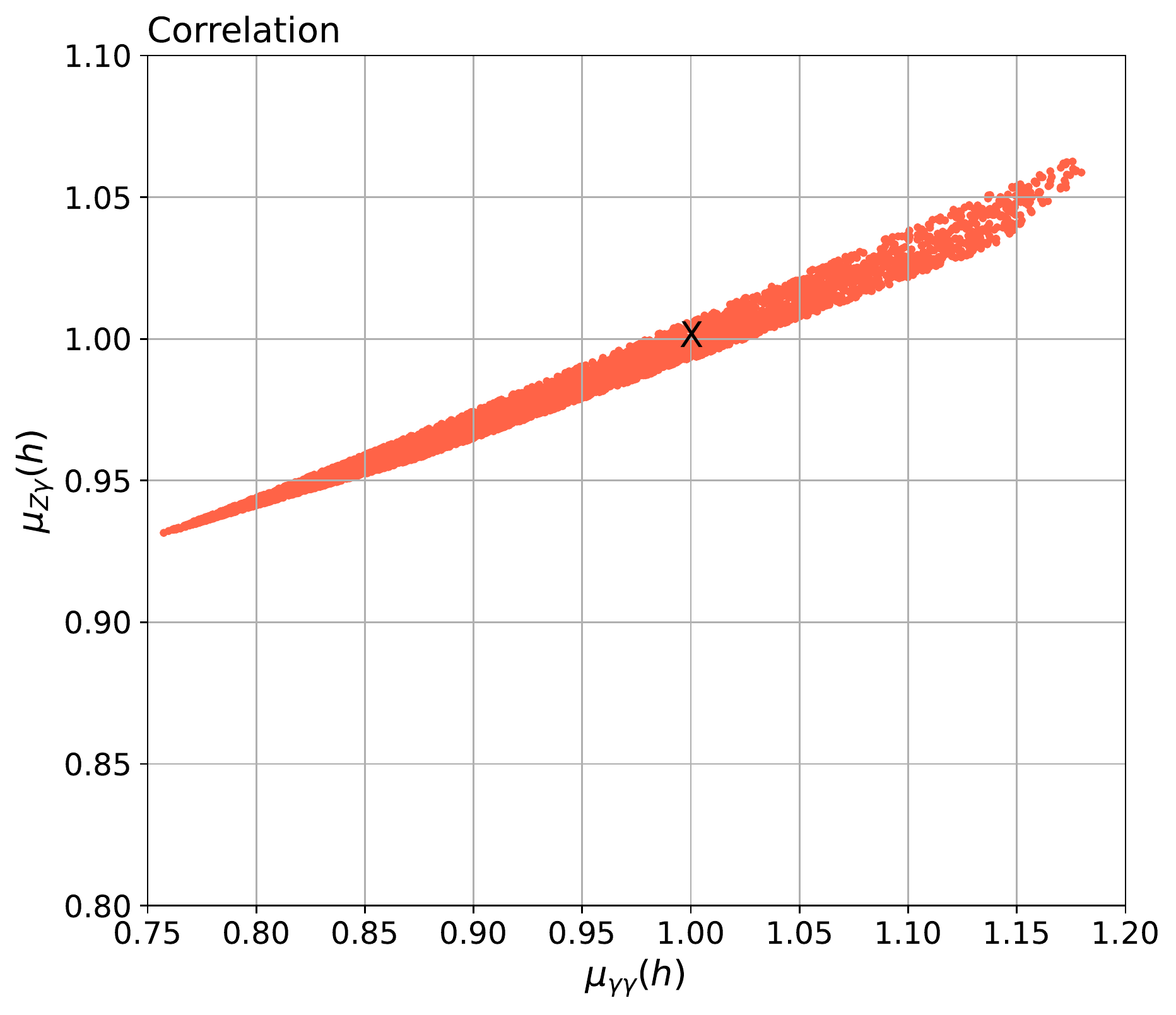}
	\end{minipage}
	\caption{$\mu_{\gamma\gamma}(h)$ and $\mu_{\gamma Z}(h)$ correlation in the HTM.}
	\label{fig:correlation-htm}
\end{figure}
\textcolor{black}{In Table. \ref{summry}. we summarize the $95\%$ CL $\mu_{ \gamma V}$ predict limits of the charged and doubly-charged Higgs boson masses based on the three non-supersymmetric models studied, taking into account the direct searches, collider searches for neutral Higgs bosons result in indirect, model-dependent limits on the charged and doubly-charged Higgs boson. In addition, we have implemented the direct signal strengths measurement at $95\%$ CL. set in Table \ref{experimental_signal_strenghts}. Hence, we found that the direct search from colliders still gives more strict limits on the charged and doubly-charged Higgs boson masses than the measured signal strengths at LHC on BSM searched.}
\begin{table}[H]
	\begin{center}
		\renewcommand{\arraystretch}{1.8}
		\adjustbox{max height=\dimexpr\textheight-10.6cm\relax,
			max width=\dimexpr\textwidth-1.8cm\relax}{
			\begin{tabular}{|c|c|c|c|c|c|c|c|c|}
				\hline
				\multicolumn{9}{|c|}{Summary of the predicted charged and doubly-charged Higgs boson masses limits from $\mu_{\gamma V}$ at $95\%$CL.﻿}\\\hline
				\multicolumn{2}{|c|}{\backslashbox{$\mu_{V \gamma  }$ limits}{modele}}&I&II&X&Y&IDM-ND&IDM-D&HTM\\\hline
				$\mu_{\gamma Z}> 1.0$&\multirow{4}{*}{$m_{H^\pm}$($m_{H^{\pm\pm}})$(GeV)}&[87.2, 113.3]&--&[98.8, 155.5]&--&NP&[127.59, 142.85]&[167.9, 195.9]([146.1, 204.2])\\
				$\mu_{\gamma Z}< 1.0$&&--&$>600$&$>95$&$>646$&--&--&[179.8, 217.9]([172.1, 244.8])\\
				$\mu_{\gamma \gamma}> 1.0$&&[86.5, 200.7]&--&[98.1, 200.1]&--&[79.46, 186.2]&[70.1, 172.1]&[167.9, 198.3]([146.1, 208.7])\\
				$\mu_{\gamma \gamma}< 1.0$&&--&$>600$&$>95$&$>646$&--&--&[178.3, 217.9]([169.0, 244.8])\\\hline
		\end{tabular}}
	\end{center}
	\caption{Predicted limits on the charged and doubly-charged Higgs boson masses from $\mu_{ \gamma V}$ at $95\%$ CL in four types of 2HDMs, IDM, and HTM. The hyphen character stands for no limits while NP is for enhancement Not possible. IDM-ND stands for IDM non-degenerate scenario while, IDM-D is for approximately degenerate scenario.}\label{summry}
\end{table}
\section{Conclusion}
\label{sec:Conclusion}
Many well-intentioned expansions of the Standard Model (SM) include a larger scalar sector with more neutral and charged Higgs bosons. It is important for us to comprehend the potential of these experiments to elucidate novel physics signals in light of the recent discovery of the SM-like 125 GeV Higgs in the ATLAS and CMS investigations. 

Recent measurements of the $\mu_{\gamma Z}$ signal strength at the LHC by the ATLAS and CMS experiments show an excess that is roughly double what the Standard Model (SM) predicted, with a significance of $2.2$ standard deviations. In this work, we have study the $\mu_{\gamma Z}$ and $\mu_{\gamma \gamma}$ excess in the following BSM frameworks : CP-conserving Two-Higgs Doublet Model (2HDMs), Inert Doublet Model (IDM) and, the Higgs triplets Model (HTM) with  hyper-charge $Y=2$, taking into account the most up-to-date available theoretical and experimental constraints.

In 2HDMs, we have found that the deviations are possible in all four types of the model and that the results from types, I (and II) are quite similar to those of types X (and Y). In type-I, the $\mu_{\gamma \gamma}(\mu_{\gamma Z})$ enhancements limited the charged Higgs boson mass $m_{H^\pm}$ to ranges from $86.3(87.2)$ \gev\ to $200.7(113.3)$ \gev, respectively. while suppression are possible irrespective of $m_{H^\pm}$. The same conclusion could be reached in type X, with a slight difference. The $\mu_{\gamma \gamma}(\mu_{\gamma Z})$ enhancement in such a case allowed a light-charged Higgs boson, ranging from  $98.1(98.8)$ \gev\ to $200.1(155.5)$ \gev, while the suppression required $m_{H^\pm} > 95$ \gev. The types II and Y allow $\mu_{\gamma \gamma}(\mu_{\gamma Z})$ enhancements everywhere, while the suppression requires a heavy-charged Higgs boson $m_{H^\pm} > 600$ \gev\ and $m_{H^\pm} > 646$ \gev,\ respectively.

In the case of the IDM model, we have demonstrated that the deviations exists for either the general or the degenerate inert states. In the former case, we have found that the  $\mu_{\gamma \gamma}$ enhancement requires the absence of the invisible decay;  i.e., $m_H$ ranges from $62.94$ \gev\ to $164.93$ \gev, and a light-charged Higgs boson; i.e., it varies  from $80$ \gev\ to $186.2$ \gev. For the $\mu_{\gamma Z }$, the enhancement is not possible. The suppression is possible in the whole parameter space for either $\mu_{\gamma \gamma}$ or $\mu_{\gamma Z}$. In the degenerate case, the $\mu_{\gamma \gamma}(\mu_{\gamma Z})$ enhancements requires light-charged Higgs and a the so called, medium dark matter candidate; $m_H$ ranges from 69.1(126.6) \gev\ to $171.1(141.85)$ \gev, while the suppression are allowed everywhere.  

For the HTM model, the deviations are possible, and either enhancement or suppression is also possible. The $\mu_{\gamma \gamma }$ and $\mu_{\gamma Z }$ enhancements occur for a light-charged and doubly-charged Higgs boson i.e., $m_{H^\pm}\in [167.9, 198.3]([167.9, 195.9])$ \gev\ and $m_{H^{\pm\pm}} \in [146.1, 208.7]([146.1, 204])$ \gev. While the suppression requires:  $m_{H^\pm} \in [179.8,217]([178.3, 217.9])$\ \gev\ and 
$m_{H^{\pm\pm}} \in [172.1, 244.8]$\\($[169.0, 244.8]$)\ \gev.
\appendix
\label{appendix}
\section{Functions $I_1(\tau,\lambda)$ and $I_2(\tau,\lambda)$}
\label{appendixA}
The explicit form exists for the functions $I_1$ and $I_2$ are in \cite{Djouadi:2005gi} as follows.
\begin{align}
&I_1(\tau,\lambda) = \frac{\tau\lambda}{2(\tau-\lambda)}+\frac{\tau^2\lambda^2}{2(\tau-\lambda)^2}\left[f(\tau)-f(\lambda) \right]+\frac{\tau^2\lambda}{(\tau-\lambda)^2}\left[g(\tau)-g(\lambda) \right],\\
&I_2(\tau, \lambda)=-\frac{\tau\lambda}{2(\tau-\lambda)}\left[f(\tau)-f(\lambda)\right]
\end{align}
where $f$ and $g$ are complex functions such that have the following properties: 
\begin{align}
&f(\tau) =
\left\{ \begin{array}{rcl}
\arcsin^{2}\sqrt{\tau} & \mbox{for}
& \tau\leq 1 \\
-\frac{1}{4}\left[\log\frac{1+\sqrt{1-\tau^{-1}}}{1-\sqrt{{1-\tau^{-1}}}}-i\pi\right]^{2} & \mbox{for} & \tau > 1
\end{array}\right.\label{fonctionf}\\
&g(\tau) =
\left\{ \begin{array}{rcl}
\sqrt{\tau^{-1}-1}\arcsin\sqrt{\tau} & \mbox{for}
& \tau\geq 1 \\
\frac{\sqrt{1-\tau^{-1}}}{2}\left[\log\frac{1+\sqrt{1-\tau^{-1}}}{1-\sqrt{{1-\tau^{-1}}}}-i\pi\right] & \mbox{for} & \tau< 1
\end{array}\right.
\end{align}
\section{The signal strengths $\mu_{\gamma \gamma}$ and $\mu_{\gamma Z }$}
\label{appendixB}
In addition, we have added experimental ATLAS and HL-LHC projections (horizontal lines) from $\mu_{\gamma Z }$ and $\mu_{\gamma \gamma}$ at $95\%$ CL in figures. \ref{fig:mu_Zga_constraints}, \ref{fig:mu_gaga_constraints}, \ref{fig:IDM_gaga_gaZ_constraints} and \ref{fig:HTM_mu_gaga_constraints}. From these figures, one can understand that the experimental signal strengths $\mu_{\gamma Z }$ and $\mu_{\gamma \gamma}$ at $95\%$ CL from ATLAS, are still far from constraining the charged Higgs boson in the four types of 2HDMs. In the case of the hoped-for HL-LHC limits, $\mu_{\gamma Z }$ remains fully consistent with the 2HDMs prediction, while $\mu_{\gamma \gamma}$ narrows the parameter space of 2HDMs types I and X without affecting the charged Higgs mass values. Therefore, bounds from $\mu_{\gamma Z }$ and $\mu_{\gamma \gamma}$ are weaker than the direct bounds at $2\sigma$ in the 2HDMs framework.}

However, in the IDM model, the $95\%$ limits from direct experimental searches of signal strengths are also given, they are still away from contributing to the charged Higgs boson mass limit. Within the hoped-for HL-LHC limits, $\mu_{\gamma Z }$ remains fully consistent with the IDM predictions in both cases studied, while $\mu_{\gamma \gamma }$ narrowed the parameters space without ruling out any values of charged Higgs mass. Therefore, bounds from $\mu_{\gamma Z }$ and $\mu_{\gamma \gamma}$ are weaker again than the direct bounds in the IDM framework.
In addition, one can see that the signal strengths $\mu_{\gamma Z }$ and $\mu_{\gamma \gamma }$ at $95\%$ CL from ATLAS, do not limit the charged and doubly-charged Higgs boson in the HTM. Within the hoped-for HL-LHC limits, the $\mu_{\gamma Z }$ still far from to be affecting the parameter space while the $\mu_{\gamma \gamma}$ rules out the regions where $m_{H^\pm}(m_{H^{\pm\pm}}) < 173.3(158.2)$ GeV and  $m_{H^\pm}(m_{H^{\pm\pm}}) > 208.8(228.3)$ GeV.
	\begin{figure}[H]
		\begin{minipage}{0.5\textwidth}
			\centering
			\includegraphics[height=4.8cm,width=6.8cm]{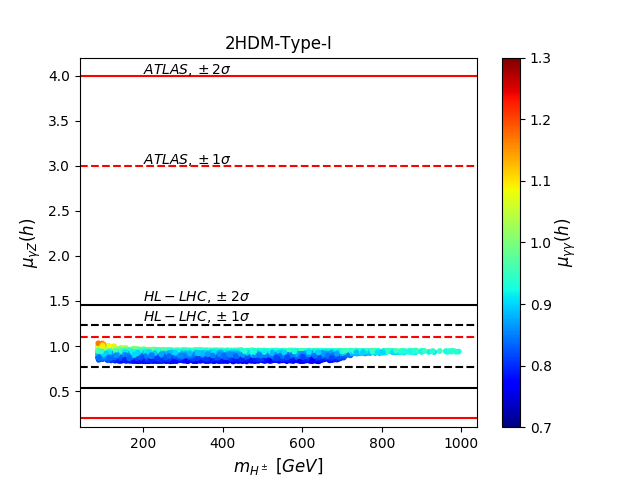}
		\end{minipage}
		\begin{minipage}{0.5\textwidth}
			\centering
			\includegraphics[height=4.8cm,width=6.8cm]{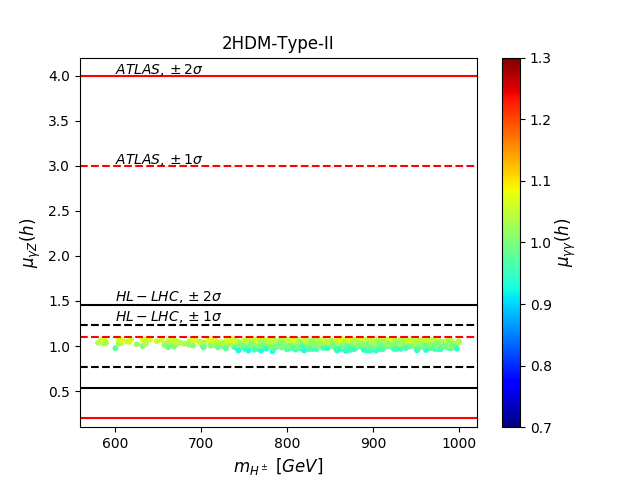}
		\end{minipage}
		\begin{minipage}{0.5\textwidth}
			\centering
			\includegraphics[height=4.8cm,width=6.8cm]{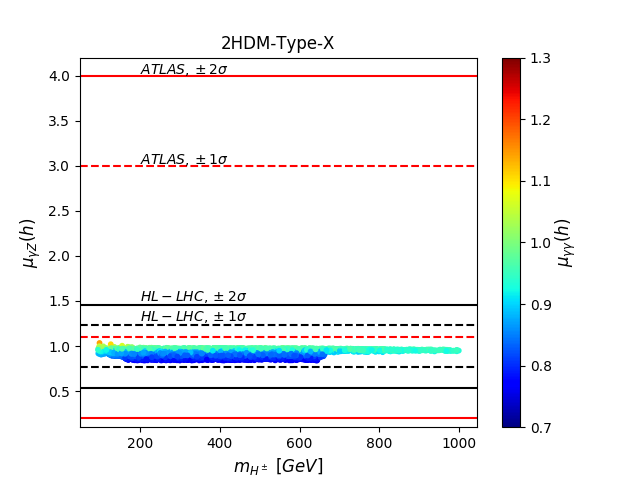}
		\end{minipage}
		\begin{minipage}{0.5\textwidth}
			\centering
			\includegraphics[height=4.8cm,width=6.8cm]{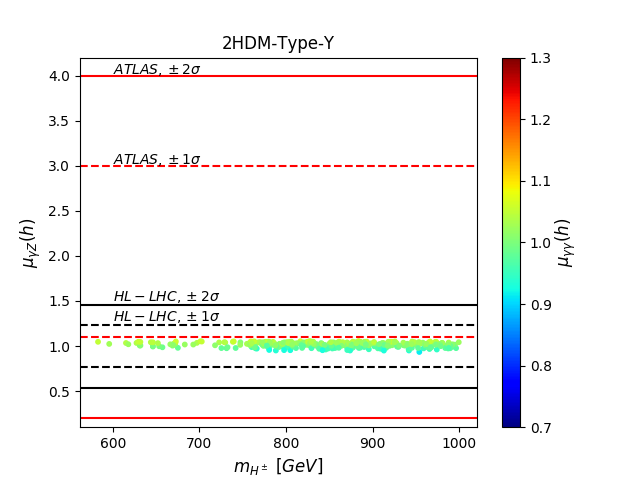}
		\end{minipage}
		\caption{The signal strength $\mu_{\gamma Z }$ at $95\%$ CL as a function of $m_{H^\pm}$ (\gev) with the color code showing $\mu_{\gamma \gamma}$ at $95\%$ CL. Types I and II (X and Y) are shown in the upper (lower) panels. The red and black horizontal lines correspond to the ATLAS and hoped-for HL-LHC experimental. }
		\label{fig:mu_Zga_constraints}
	\end{figure}
	\begin{figure}[H]
		\begin{minipage}{0.5\textwidth}
			\centering
			\includegraphics[height=4.8cm,width=6.8cm]{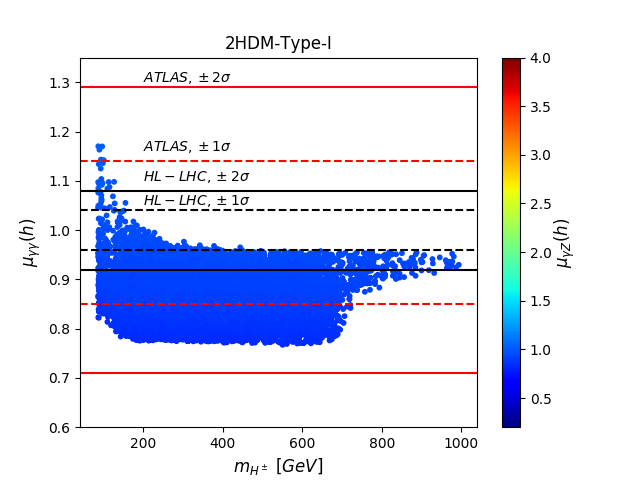}
		\end{minipage}
		\begin{minipage}{0.5\textwidth}
			\centering
			\includegraphics[height=4.8cm,width=6.8cm]{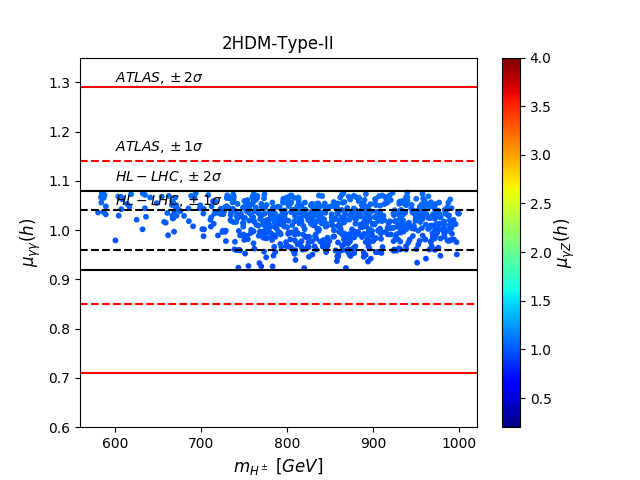}
		\end{minipage}
		\begin{minipage}{0.5\textwidth}
			\centering
			\includegraphics[height=4.8cm,width=6.8cm]{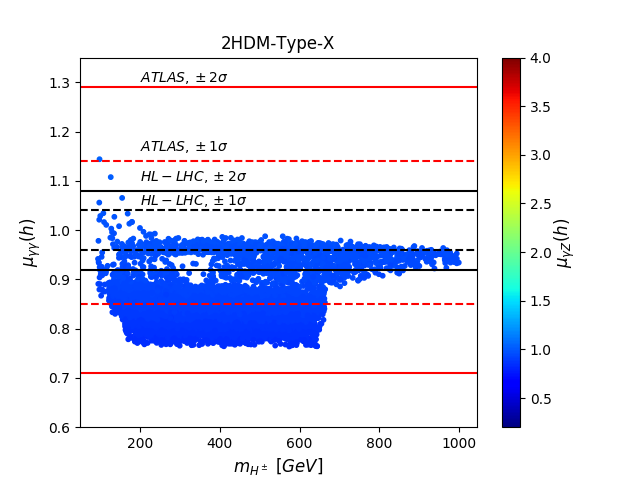}
		\end{minipage}
		\begin{minipage}{0.5\textwidth}
			\centering
			\includegraphics[height=4.8cm,width=6.8cm]{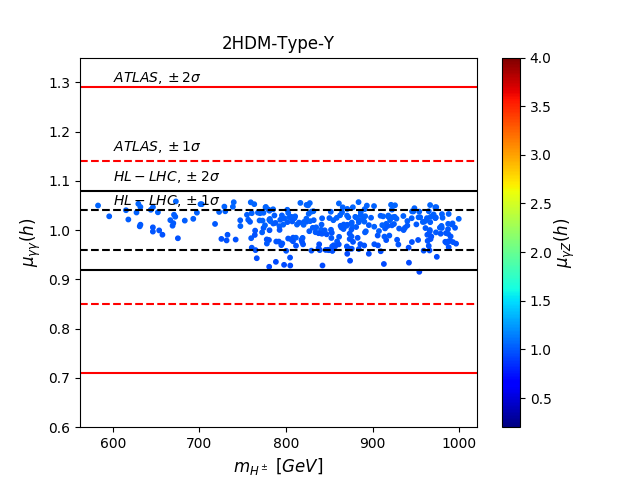}
		\end{minipage}
		\caption{The same as Fig. \ref{fig:mu_Zga_constraints} for $\mu_{\gamma\gamma}$ while $\mu_{\gamma Z }$ is in the palette.}
		\label{fig:mu_gaga_constraints}
	\end{figure}
	\begin{figure}[H]
		\begin{minipage}{0.5\textwidth}
			\centering
			\includegraphics[height=4.8cm,width=6.8cm]{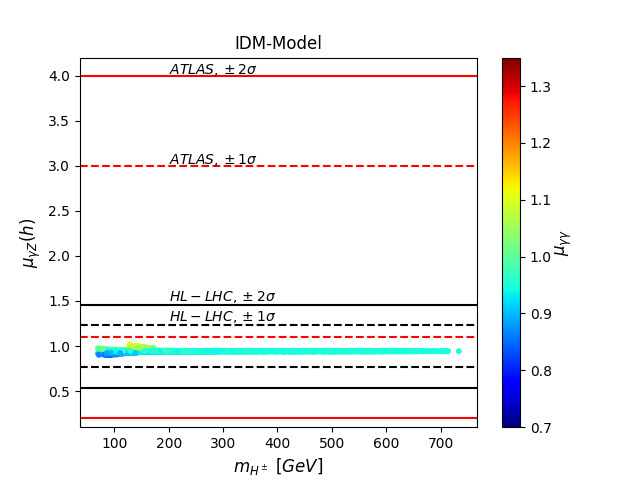}
		\end{minipage}
		\begin{minipage}{0.5\textwidth}
			\centering
			\includegraphics[height=4.8cm,width=6.8cm]{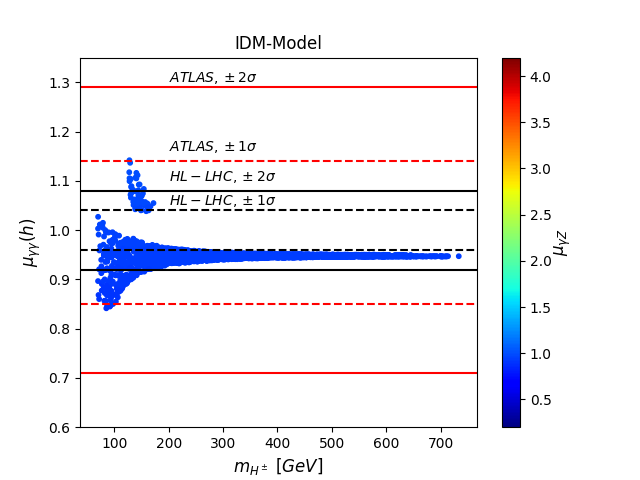}
		\end{minipage}
		\begin{minipage}{0.5\textwidth}
			\centering
			\includegraphics[height=4.8cm,width=6.8cm]{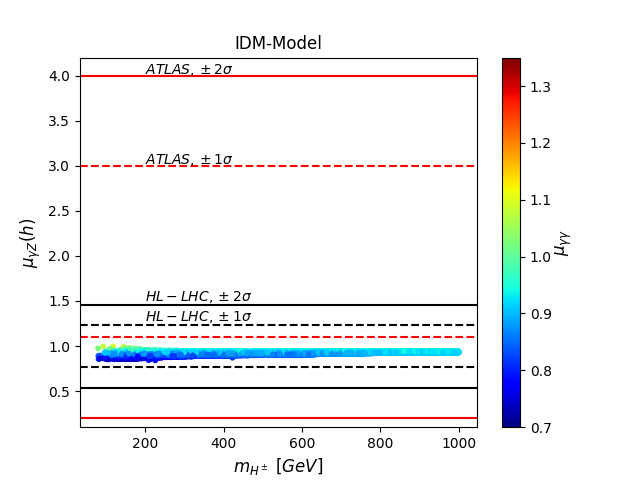}
		\end{minipage}
		\begin{minipage}{0.5\textwidth}
			\centering
			\includegraphics[height=4.8cm,width=6.8cm]{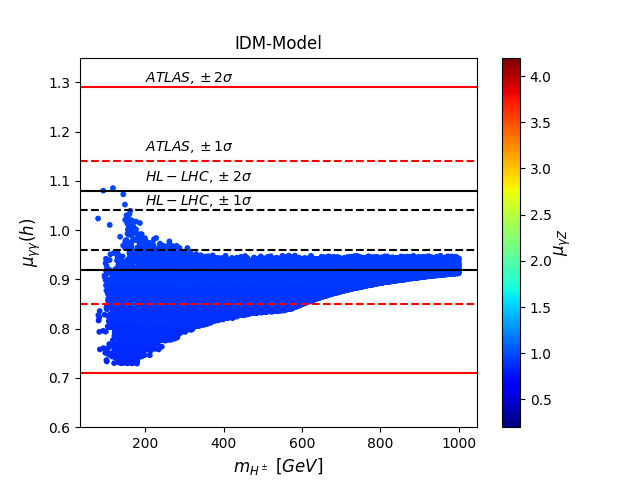}
		\end{minipage}
		\caption{The same as Fig. \ref{fig:mu_Zga_constraints} (and \ref{fig:mu_gaga_constraints}) for the IDM approximate degenerate (upper) and non-degenerate (lower) panels. }
		\label{fig:IDM_gaga_gaZ_constraints}
	\end{figure}
	\begin{figure}[H]
		\begin{minipage}{0.5\textwidth}
			\centering
			\includegraphics[height=4.8cm,width=6.8cm]{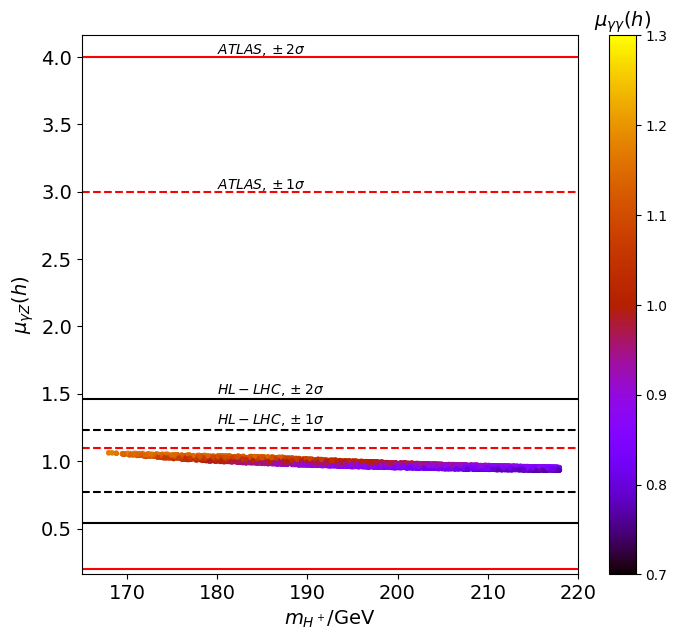}
		\end{minipage}
		\begin{minipage}{0.5\textwidth}
			\centering
			\includegraphics[height=4.8cm,width=6.8cm]{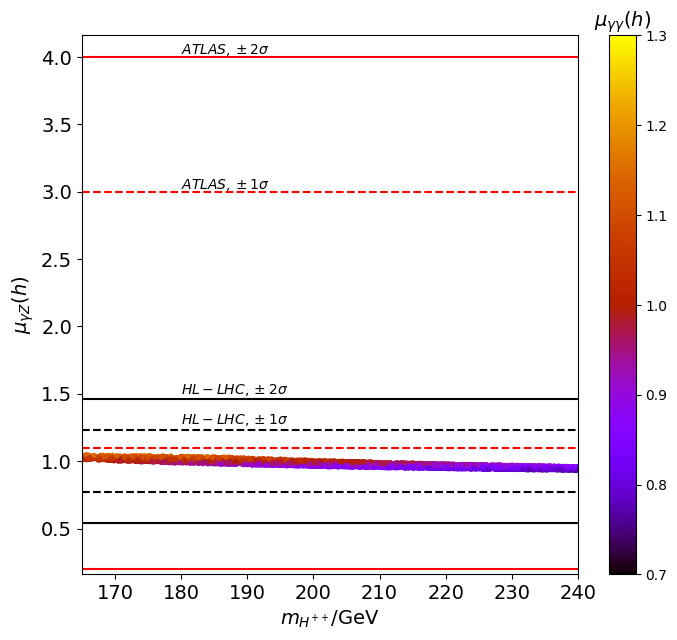}
		\end{minipage}
		\begin{minipage}{0.5\textwidth}
			\centering
			\includegraphics[height=4.8cm,width=6.8cm]{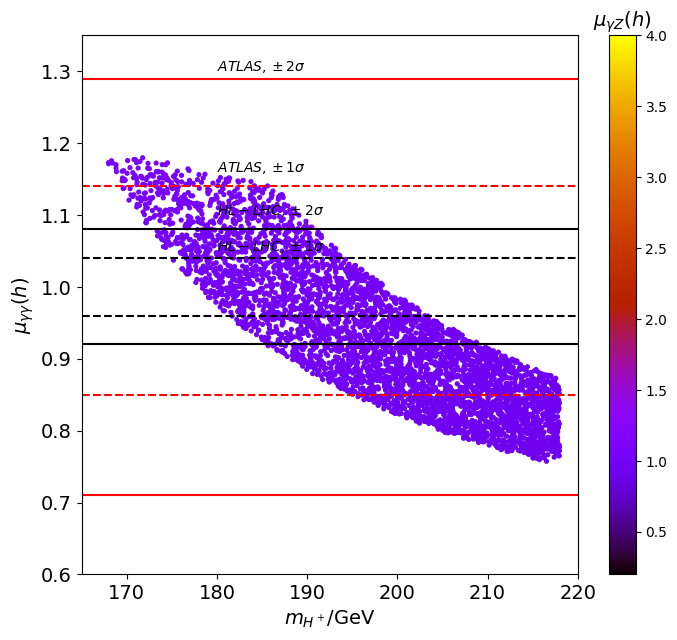}
		\end{minipage}
		\begin{minipage}{0.5\textwidth}
			\centering
			\includegraphics[height=4.8cm,width=6.8cm]{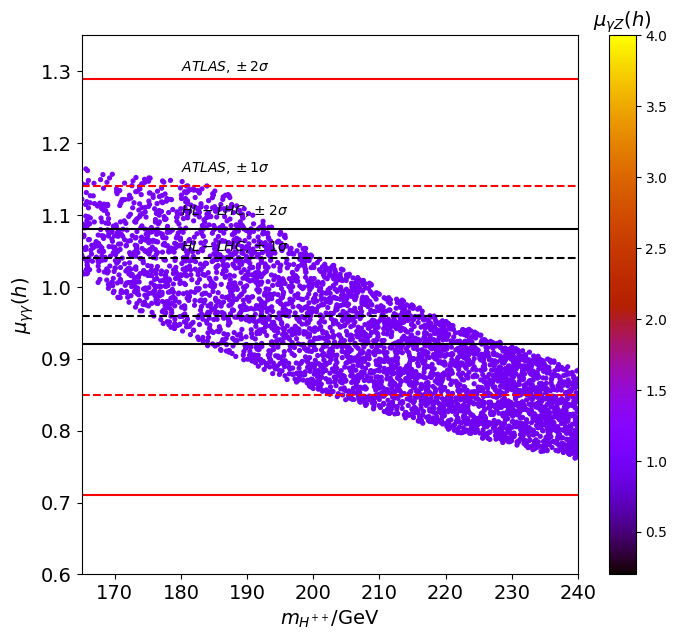}
		\end{minipage}
	\caption{The same as Fig. \ref{fig:mu_Zga_constraints} (and \ref{fig:mu_gaga_constraints}) for the HTM but, this time as a function of $m_{H^\pm}$ (left) and $m_{H^{\pm\pm}}$ (right) panels. }
		\label{fig:HTM_mu_gaga_constraints}
	\end{figure}
\section*{Acknowledgments}
This work is supported by the Moroccan Ministry of Higher Education and Scientific Research
MESRSFC and CNRST: Projet PPR/2015/6. MO is grateful for the technical support of CNRST/HPC-MARWAN.
\bibliographystyle{JHEP}
\bibliography{references}
\end{document}